\RequirePackage[T1]{fontenc}
\documentclass[12pt]{article}

\usepackage[height=8.85in,width=6.45in]{geometry}

\usepackage[utf8]{inputenc}
\usepackage{amsmath}
\usepackage{amssymb}
\usepackage{cancel}
\usepackage{mathtools}
\numberwithin{equation}{section}
\usepackage{slashed}
\usepackage{braket}
\usepackage{enumitem}
\usepackage[svgnames,dvipsnames]{xcolor}
\usepackage[colorlinks,citecolor=DarkGreen,linkcolor=FireBrick,urlcolor=FireBrick,linktocpage,unicode,psdextra]{hyperref}
\usepackage{cite}
\usepackage{booktabs}
\usepackage{multirow}
\usepackage{graphicx}
\usepackage{longtable}
\usepackage[bottom]{footmisc}

\usepackage{times}
\usepackage{courier}
\usepackage{bm}
\usepackage{subfig}
\usepackage{adjustbox}

\usepackage{colortbl}
\usepackage{mdframed}

\renewenvironment{figure}[1][]{
  \begin{originalfigure}[#1]
    \begin{mdframed}[linecolor=black!0,backgroundcolor=white]
}{
    \end{mdframed}
  \end{originalfigure}
}

\usepackage{amsthm}
\theoremstyle{plain}

\numberwithin{them}{section}

\theoremstyle{remark}

\def\ZZ{\mathbb{Z}}

\def\p{\partial}
\newcommand{\bea}{\begin{eqnarray}}
\newcommand{\eea}{\end{eqnarray}}
\def\no{\nonumber}
\def\m{\mu}

\def\cM{\mathcal{M}}

\def\cL{\mathcal{L}}
\def\RR{\mathbb{R}}

\def\half{{1\over 2}}

\usepackage{dsfont}

\def\T{{\mathsf T}}
\def\F{\mathsf{F}}

\def\Ls{L_{\rm SUSY}}
\newcommand{\minus}{-}
\def\Ln{L_{\cancel{\rm SUSY}}}

\usepackage{tikz}
\usetikzlibrary{decorations.markings}
\usetikzlibrary{arrows.meta,positioning}

\begin{document}

\begin{titlepage}

\begin{flushright}
KYUSHU-HET-365
\end{flushright}

\vskip 3cm

\begin{center}

{\Large \bfseries Maximal Enhancements in Eight-Dimensional Non-Supersymmetric Heterotic Strings}

\vskip 2cm

Yamato Honda$^1$, 
Justin Kaidi$^{1,2,3}$,
Yuichi Koga$^{2}$,
Yuefeng Liu$^{2}$
\vskip 1cm

\begin{tabular}{c p{0.78\textwidth}}
1 & Department of Physics, Kyushu University, Fukuoka 819-0395, Japan \\
2 & Institute for Advanced Study, Kyushu University, Fukuoka 819-0395, Japan\\
3 & Quantum and Spacetime Research Institute (QuaSR), Kyushu University, Fukuoka 819-0395, Japan
\end{tabular}

\vskip 2cm

\end{center}

\noindent

We classify maximally enhanced, rank-preserving non-supersymmetric heterotic strings in eight dimensions by studying orbifolds of supersymmetric heterotic strings on $T^2$.  We first review the construction in nine dimensions, where it reproduces the $95$ maximally semisimple enhancement points obtained from previous extended-Dynkin analyses.  We then start from the maximally enhanced supersymmetric Narain points in eight dimensions and use Kac's theorem to enumerate candidate order-two inner actions, retaining only those that lift to consistent Narain-lattice shifts.  This gives $1210$ maximal enhancement points, of which $71$ are tachyon-free, and $25$ have neither tachyons nor shifted-sector massless scalars.  For each entry we determine the gauge algebra and the low-lying scalar and fermion spectra, and for the tachyon-free cases we evaluate the one-loop cosmological constant.  For the 25 subcases without shifted-sector massless scalars, we further compute the Hessian of the one-loop potential and test the refined de Sitter swampland conjecture. 
\end{titlepage}

\setcounter{tocdepth}{2}
\tableofcontents

\section{Introduction}
\label{sec:introduction}

Non-supersymmetric string vacua provide a controlled setting in which to study phenomena that are hidden by spacetime supersymmetry: perturbative tachyons, one-loop vacuum energies, scalar instabilities, and the relation between enhanced gauge symmetry and extrema of the quantum effective potential.  Heterotic strings are particularly useful for this purpose.  Toroidal compactifications admit an exact Narain-lattice description \cite{Narain:1985jj,Narain:1986amc}, and the extra left-moving current algebra gives a simple understanding of non-Abelian gauge enhancement.  At the same time, non-supersymmetric heterotic strings exhibit many of the interesting phenomena expected in non-supersymmetric quantum gravity and provide a useful framework for exploring phenomenological questions associated with supersymmetry breaking, see e.g. \cite{Abel:2015oxa,Funakoshi:2025lxs,Escalante-Notario:2025hvn,Detraux:2026gjk}.

There are two complementary approaches to obtaining rank-preserving non-supersymmetric heterotic compactifications.  One approach is to begin with a ten-dimensional non-supersymmetric heterotic string and study its toroidal moduli space directly; this is the viewpoint of the classic analyses in \cite{Nair:1986zn,Ginsparg:1986wr}, revisited recently in  e.g. \cite{Fraiman:2023cpa}. In particular, in the case of compactifications to nine dimensions, \cite{Fraiman:2023cpa} constructed an extended Dynkin diagram that encodes all walls in the fundamental domain of the nine-dimensional moduli space.  In the conventions of that reference, there are $107$ reflection-fixed maximal-enhancement points. However, twelve retain a $U(1)$ factor, and with the stricter convention adopted in this paper, in which ``maximally enhanced'' means maximally \emph{semisimple} on the left-moving side, the corresponding number is $95$.\footnote{Note that by this definition, there are only five maximally enhanced non-supersymmetric theories in ten dimensions---the $\mathfrak{u}(16)$ heterotic string is not included due to the presence of the $\mathfrak{u}(1)$ factor. See Appendix \ref{app:10d} for further details.} Eight theories are found to be tachyon-free.

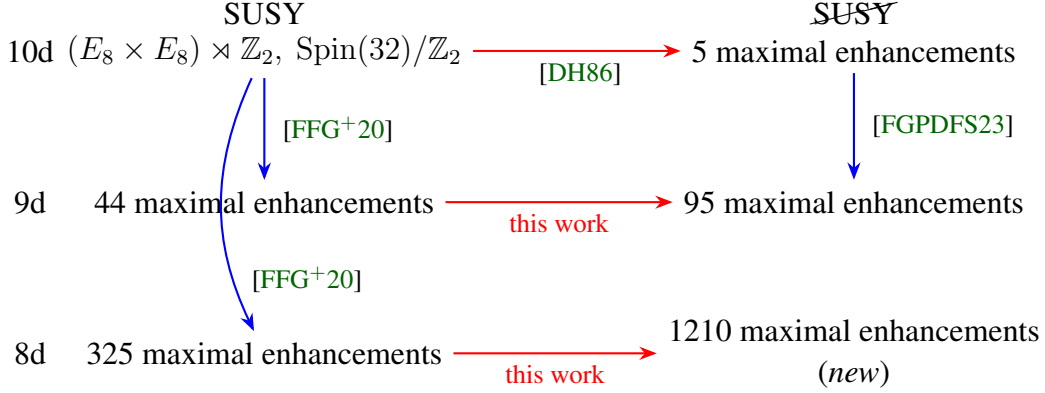
\begin{figure}[tb]
\centering

\begin{tikzpicture}[
  node distance=2.5cm and 3cm,
  every node/.style={align=center},
  arrow/.style={-{Stealth}, thick},
  dashedarrow/.style={-{Stealth}, thick, dashed},
  labelstyle/.style={font=\footnotesize}
]

\node (SUSY) at (-3.4,4.5) {SUSY};

\node (10dleft) at (-3.4,4)
  {$(E_8 \times E_8)\rtimes \ZZ_2,\ \mathrm{Spin}(32)/\mathbb{Z}_2$};

\node (9dleft) at (-3.4,2)
  {44 maximal enhancements};

\node (8dleft) at (-3.4,0)
  {325 maximal enhancements};

\node (non-SUSY) at (4.4,4.5)
  {\cancel{SUSY}};

\node (10dright) at (4.4,4)
  {5 maximal enhancements};

\node (9dright) at (4.4,2)
  {95 maximal enhancements};

\node (8dright) at (4.4,0)
  {1210 maximal enhancements\\(\textit{new})};

\node (10d) at (-6.5,4) {10d};
\node (9d)  at (-6.5,2) {9d};
\node (8d)  at (-6.5,0) {8d};

  \draw[arrow,draw=red]
  (10dleft) -- (10dright)
  node[midway, below, labelstyle]
  {\cite{Dixon:1986iz}};

\draw[arrow,draw=blue]
  (10dleft) -- (9dleft)
  node[midway, right, labelstyle]
  {\cite{Font:2020rsk}};

\draw[arrow, draw=red]
  (9dleft) -- (9dright)
  node[midway, below, labelstyle, text=red]
  {this work};

\draw[arrow,draw=blue]
  (10dleft)
  to[bend right=25]
  node[pos=0.8, right, labelstyle]
  {\cite{Font:2020rsk}}
  (8dleft);

\draw[arrow, draw=red]
  (8dleft) -- (8dright)
  node[midway, below, labelstyle, text=red]
  {this work};

\draw[arrow,draw=blue]
  (10dright) -- (9dright)
  node[midway, right, labelstyle]
  {\cite{Fraiman:2023cpa}};

\end{tikzpicture}

\caption{A schematic overview of the connections among rank-preserving heterotic theories in ten, nine, and eight dimensions. Blue lines denote toroidal compactification, while red lines denote $\ZZ_2$ orbifolds.}
\label{fig:connection}

\end{figure}

The second approach is to start with a supersymmetric heterotic compactification and quotient by an order-two translation of the internal Narain theory combined with spacetime fermion parity. In a geometric duality frame, this is a freely acting orbifold; after passing to the quotient circle it is equivalently a Scherk-Schwarz-like twisted compactification \cite{Scherk:1978ta,Rohm:1983aq,Kounnas:1989dk}. 
To understand this approach better, recall that a ten-dimensional heterotic string is specified by a chiral fermionic CFT of central charge sixteen. Every such theory is obtained by fermionizing a bosonic chiral CFT \cite{BoyleSmith:2023xkd,Hohn:2023auw,Rayhaun:2023pgc}, and since the only bosonic theories of central charge sixteen are those associated with the supersymmetric $(E_8\times E_8)\rtimes \ZZ_2$ and $\mathrm{Spin}(32)/\ZZ_2$ heterotic strings, every ten-dimensional non-supersymmetric heterotic string is obtained by gauging an order-two symmetry of a supersymmetric theory together with spacetime fermion parity \cite{Dixon:1986iz}. For the rank-preserving theories relevant to the current work, this order-two symmetry is an inner automorphism that preserves all Cartan currents and is therefore represented, in the lattice description, by an order-two shift. Since the Narain current-current deformations are invariant under this shift, compactification and deformation may be performed either before or after the gauging. Consequently, every theory in the standard rank-preserving toroidal component is an order-two quotient of a supersymmetric toroidal theory in the same spacetime dimension.

It remains to determine which supersymmetric theories can give rise to a maximally enhanced non-supersymmetric theory. For the shift gaugings under consideration in this work, every left-moving gauge current of the non-supersymmetric daughter is inherited from the supersymmetric parent. The gauge algebra of the daughter is therefore a subalgebra of that of the parent. If the daughter has the maximum possible rank, then the parent must have at least the same rank; since no supersymmetric heterotic theory in that dimension can have a larger rank, the parent must itself be maximally enhanced. It follows that starting from the complete list of maximally enhanced supersymmetric toroidal theories and considering all consistent rank-preserving order-two shifts exhausts the maximally enhanced non-supersymmetric theories in this component.\footnote{This conclusion does not apply to rank-reduced CHL-like components or to more general asymmetric constructions. The classification for the rank-reduced (non-)supersymmetric branches can be found in e.g. \cite{Font:2021uyw,Nakajima:2023zsh,DeFreitas:2024ztt,Hamada:2024cdd,Hamada:2025cpe,Fraiman:2025yrx}.}

In \cite{Font:2020rsk}, the complete list of maximally enhanced supersymmetric theories was given for toroidal compactifications to nine and eight dimensions. Thus, in order to classify maximally enhanced non-supersymmetric theories in these dimensions, it suffices to classify all possible order-two inner automorphisms in the known supersymmetric theories.
In the current work, we perform this classification by making use of Kac's theorem, which converts the problem into a finite calculation on affine Dynkin diagrams \cite{kac1990infinite}. Having done so, we enumerate all maximally enhanced non-supersymmetric theories in dimensions nine and eight, and study their various properties. The basic chain of ideas is as shown in Figure \ref{fig:connection}.  In particular, the nine-dimensional results match with those obtained in \cite{Fraiman:2023cpa} and serve as a useful check of our methodology, while the eight-dimensional results are new.

To summarize the new results, our construction yields a total of $1210$ eight-dimensional entries, of which $71$ are tachyon-free.  In particular, $25$ entries have no shifted-sector massless scalars, while $46$ lie on a knife edge according to the criterion of \cite{Ginsparg:1986wr}.\footnote{A ``knife edge'' is a point where shifted sectors contain additional massless scalars beyond the universal massless scalars associated with the classical Narain moduli.} The sign of the one-loop cosmological constant is positive for $68$ entries and negative for three in the normalization used below.  Neither absence of a tree-level tachyon nor absence of a knife-edge scalar implies perturbative stability: indeed, we compute the Hessian of the one-loop potential, finding that one vacuum is a local maximum, while the others are saddle points.
For ease of reference, Table \ref{tab:intro8d} lists the $25$ scalar-free tachyon-free entries.\footnote{Throughout this work, we use the term ``scalar-free'' to describe theories that are not located at knife-edge points in moduli space. This does not exclude a massless dilaton or massless Narain moduli. }    The parent algebra and a complete representative of the Narain shift are given in Appendix \ref{app:8d}; Section \ref{sec:eightdimensions} explains the conventions used in both tables.

Our new eight-dimensional results can also be used to check the status of the refined de Sitter swampland conjecture \cite{Obied:2018sgi,Garg:2018reu,Ooguri:2018wrx}. The highly stringy enhanced-symmetry points studied here lie outside the parametrically controlled regime of low-energy effective field theory and therefore provide an interesting setting in which to test the conjecture directly using the string one-loop potential. We find that the relevant dimensionless Hessian ratio ${\min} \left(g^{ac}\nabla_c \nabla_b V\right)/V$ is of order unity,  as expected from the refined de Sitter conjecture.

\begingroup
\small
\begin{longtable}{@{}c c p{0.07\textwidth} c c c@{}}
\toprule
$k$ & $\Ln$ & $N_0^{\rm f}$ & $\Lambda^{(8)}\times10^6$ & ${\min} \left(g^{ac}\nabla_c \nabla_b \Lambda^{(8)}\right)/\Lambda^{(8)}$ \\
\midrule
\endfirsthead
\toprule
 $k$ & $\Ln$ & $N_0^{\rm f}$ & $\Lambda^{(8)}\times10^6$ & ${\min} \left(g^{ac}\nabla_c \nabla_b V\right)/V$ \\
\midrule
\endhead
\midrule
\multicolumn{5}{r}{\emph{continued on next page}}\\
\endfoot
\hline\\
 \caption{The scalar-free, tachyon-free theories identified in eight dimensions. Further information is given in Table \ref{tab:8d} and Appendix \ref{app:8d}. The label $k$ denotes the parent supersymmetric eight-dimensional theory following the notation of \cite{Font:2020rsk}, while $\Ln$ is the gauge algebra of the non-supersymmetric theory, $N_0^{\rm f}$ is the number of massless fermions, and $\Lambda^{(8)}$ is the eight-dimensional cosmological constant.
}
\label{tab:intro8d}
\endlastfoot
 $1$ & $\scriptstyle 6A_{3}$ & 108 & 106.82 & $-0.67$ \\
 $25$ & $\scriptstyle 4A_{1} + 2A_{7}$ & 156 & 106.82 & $-2.95$ \\
 $28$ & $\scriptstyle 2A_{1} + 3A_{3} + A_{7}$ & 130 & 106.30 & $-3.80$ \\
 $113$ & $\scriptstyle 2A_{4} + 2D_{5}$ & 100 & 145.97 & $-1.56$ \\
 $121$ & $\scriptstyle A_{1} + A_{7} + 2D_{5}$ & 170 & 147.45 & $-0.69$ \\
 $122$ & $\scriptstyle A_{1} + A_{2} + A_{3} + A_{7} + D_{5}$ & 130 & 130.69 & $-1.00$ \\
 $123$ & $\scriptstyle 2A_{1} + A_{4} + A_{7} + D_{5}$ & 110 & 120.48 & $-2.20$ \\
 $124$ & $\scriptstyle A_{8} + 2D_{5}$ & 100 & 111.94 & $-2.11$ \\
 $137$ & $\scriptstyle 2A_{3} + 2D_{6}$ & 180 & 106.82 & $-2.95$ \\
 $154$ & $\scriptstyle A_{7} + D_{5} + D_{6}$ & 190 & 106.30 & $-3.80$ \\
 $177$ & $\scriptstyle 2D_{5} + D_{8}$ & 228 & 106.82 & $-2.95$ \\
 $179$ & $\scriptstyle 2D_{9}$ & 324 & 106.82 & $-2.95$ \\
 $187$ & $\scriptstyle A_{4} + D_{5} + D_{9}$ & 180 & 117.13 & $-4.63$ \\
 $299$ & $\scriptstyle 2A_{1} + 2A_{4} + D_{8}$ & 192 & 142.49 & $-2.41$ \\
$199$ & $\scriptstyle A_{1} + A_{2} + A_{4} + D_{5} + D_{6}$ & 184 & 155.53 & $-1.62$ \\
 $199$ & $\scriptstyle A_{1} + A_{2} + A_{3} + A_{4} + D_{8}$ & 224 & 154.51 & $-1.74$ \\
 $256$ & $\scriptstyle 2A_{1} + 2A_{2} + 2D_{6}$ & 272 & 166.26 & $-0.91$ \\
 $316$ & $\scriptstyle 2A_{1} + 2A_{2} + D_{4} + D_{8}$ & 288 & 166.26 & $-0.91$ \\
 $314$ & $\scriptstyle A_{1} + A_{4} + D_{5} + D_{8}$ & 288 & 175.58 & $-1.62$ \\
 $315$ & $\scriptstyle A_{5} + D_{5} + D_{8}$ & 288 & 147.19 & $-1.73$ \\
$316$ & $\scriptstyle 2A_{2} + D_{6} + D_{8}$ & 320 & 166.26 & $-0.91$ \\
 $211$ & $\scriptstyle 2A_{1} + A_{2} + D_{6} + D_{8}$ & 384 & 187.77 & $-0.82$ \\
 $212$ & $\scriptstyle A_{1} + A_{3} + D_{6} + D_{8}$ & 384 & 165.36 & $-0.92$ \\
 $296$ & $\scriptstyle 2A_{1} + 2D_{8}$ & 512 & 211.74 & $-0.73$ \\
 $297$ & $\scriptstyle A_{2} + 2D_{8}$ & 512 & 196.54 & $-0.79$ \\
\end{longtable}
\endgroup

\paragraph{Organization.}
We begin in Section \ref{sec:review} by reviewing the Narain description of supersymmetric heterotic compactifications, and introduce the relevant order-two orbifolds.  We then explain how Kac's theorem enumerates the possible inner actions at a maximally enhanced point, how these Lie algebra data are lifted to genuine vectors of the Narain lattice, and how the resulting orbifold partition function is assembled.  The section closes by describing a finite search algorithm for tachyons and massless states.  Section \ref{sec:nine-dimensions} then applies this machinery in nine dimensions and compares it with the extended-Dynkin classification of \cite{Fraiman:2023cpa}, while Section \ref{sec:eightdimensions} presents the eight-dimensional classification and explains the data in Table \ref{tab:intro8d}.  

To keep the discussion self-contained, in Appendix \ref{app:10d} we review the standard orbifold approach for ten-dimensional non-supersymmetric heterotic strings, while in Appendix \ref{app:modulispace} we review some details about the metric on moduli space of toroidal compactifications (necessary for computing the Hessian).  Appendix \ref{app:8d} collects details on the various eight- and nine-dimensional theories identified in this work.

Finally, a complete list of all 1210 maximally enhanced eight-dimensional theories is provided as an ancillary file with the arXiv submission. 

\section{From supersymmetric to non-supersymmetric strings }
\label{sec:review}

We begin by reviewing the Narain description of toroidal heterotic compactifications \cite{Narain:1985jj,Narain:1986amc}, together with the orbifold construction of non-supersymmetric theories \cite{Dixon:1986iz,Ginsparg:1986wr}.
Much of this content is standard and can be skipped by the familiar reader, though the bounds discussed in Section \ref{sec:light-states} may be slightly unfamiliar. 

\subsection{Narain compactification and order-two symmetries}
The starting point is a supersymmetric heterotic string on $T^d$.  Its internal momenta form the even self-dual Narain lattice $\Gamma_{16+d,d}$, whose elements may be written as
\bea
\label{eq:Narainvector}
P=(P_L;P_R)=(\ell_L,p_L;p_R)~.
\eea
The split $P_L=(\ell_L,p_L)$ separates the $16$ chiral bosons that realize the ten-dimensional gauge lattice from the $d$ left-moving bosons associated with the compactification torus.  The right-moving supersymmetric sector has only the geometric momentum $p_R\in\mathbb R^d$.  This decomposition depends on the chosen polarization of the Narain lattice, whereas the full Narain lattice and Lorentzian norm are invariant under $O(16+d,d;\mathbb Z)$.

In terms of Kaluza-Klein momenta $n_i$, winding numbers $w^i$, and a gauge-lattice vector $\Pi$, the momenta can be written as (see e.g. \cite{Giveon:1994fu})
\bea
\label{eq:ellppdef}
\ell_L &=& \Pi+A_iw^i~,
\no\\
p_{L,a}&=&\frac1{\sqrt2}e^{*i}_a\left[n_i+(2G_{ij}-E_{ij})w^j-\Pi\cdot A_i\right]~,
\no\\
p_{R,a}&=&\frac1{\sqrt2}e^{*i}_a\left[n_i-E_{ij}w^j-\Pi\cdot A_i\right]~.
\eea
Here $e^a_i$ is a vielbein for $G_{ij}$ and $e^{*i}_a$ is its dual,
\bea
 e^a_ie^a_j=G_{ij}~, \hspace{0.4 in} e^a_i e^{*j}_a=\delta_i{}^j~, \hspace{0.4 in} e^{*i}_ae^{*j}_a=(G^{-1})^{ij}~,
\eea
while
\bea
E_{ij}:=G_{ij}+B_{ij}+\frac12A_i\cdot A_j~.
\eea
The vector $\Pi$ lies in $\Gamma_{16}=\Gamma_{E_8}\oplus\Gamma_{E_8}$ or $\Gamma_{\mathrm{Spin}(32)/\mathbb Z_2}$.  The Lorentzian pairing is independent of the continuous moduli,
\bea
\label{eq:Pinnerprod}
P_1\eta P_2
=P_{L1}\cdot P_{L2}-P_{R1}\cdot P_{R2}
=\Pi_1\cdot\Pi_2+w_1^in_{2i}+n_{1i}w_2^i~,
\eea
with $\eta:=\operatorname{diag}(\mathds 1_{16+d},-\mathds 1_d)$.

We now choose a half-lattice translation $\T_\delta$ acting by
\bea
\label{eq:Tdeltadef}
\T_\delta|P\rangle=e^{2\pi iP\cdot\delta}|P\rangle~, \hspace{0.4 in} 2\delta\in\Gamma_{16+d,d}~.
\eea
The condition $2\delta\in\Gamma_{16+d,d}$ makes this an order-two operation, since $2P\cdot\delta\in\mathbb Z$ for every $P\in\Gamma_{16+d,d}$.  The non-supersymmetric theory is then obtained by gauging
\bea
\label{eq:gdeltadef}
g_\delta:=(-1)^\F\T_\delta~,
\eea
where $\F$ is spacetime fermion number.  When the lattice translation can be represented as a geometric half-period shift, the quotient is freely acting.  In this case there is an equivalent interpretation as a Scherk-Schwarz compactification---namely, fields are periodic up to $(-1)^\F$ and the accompanying internal translation.  However, in a different duality frame the same $\T_\delta$ can mix geometric, T-dual, and gauge directions and need not look geometric \cite{Scherk:1978ta,Rohm:1983aq,Kounnas:1989dk,Narain:1986qm}.

The consistency conditions that must be satisfied by $\delta$ are then as follows,
\bea
\label{eq:deltaconstraints}
2\delta\in\Gamma_{16+d,d}~,
\qquad \delta^2\in\mathbb Z~.
\eea
The second condition is the order-two level-matching, or anomaly-cancellation, condition.  Its origin is made explicit in Section \ref{sec:orbifold}.

\subsection{Kac's theorem}
\label{sec:Kactheorem}
At a maximally enhanced supersymmetric point, additional massless gauge bosons arise from vectors satisfying
\bea
\label{eq:gaugeenhancement}
P_L^2=2~,
\qquad P_R=0~.
\eea
They form the root system $\Delta(\Ls)$ of a simply-laced algebra of rank $16+d$.  We use Kac's theorem to enumerate the possible order-two inner actions on this root system \cite{kac1990infinite}.  The full theorem also includes outer automorphisms; restricting to inner actions ensures that all Cartan currents survive and therefore restricts us to rank-preserving branches.

Let $L$ be a simple Lie algebra of rank $r$, with simple roots $\alpha_i$, fundamental weights $\omega_i$, and lowest root
\bea
\alpha_0:=-\sum_{i=1}^r\kappa_i\alpha_i~,  \hspace{0.4 in}  \kappa_0:=1~.
\eea
The $\kappa_i$ are known as Kac marks.  An inner automorphism whose order divides $N$ can be represented by non-negative affine labels $(s_0,s_1,\ldots,s_r)$ satisfying
\bea
\label{eq:Kacequations}
N=\sum_{i=0}^rs_i\kappa_i~, \hspace{0.4 in} \boldsymbol\delta_L=\frac1N\sum_{i=1}^rs_i\omega_i~.
\eea
The automorphism acts on the root generators as $E_\alpha\mapsto e^{2\pi i\alpha\cdot\boldsymbol\delta_L}E_\alpha$, and hence the surviving roots are
\bea
\label{eq:DeltaLndef}
\Delta(\Ln)=\{\alpha\in\Delta(\Ls)\mid \alpha\cdot\boldsymbol\delta_L\in\mathbb Z\}~.
\eea
For primitive Kac data with $p$ nonzero labels, the fixed algebra contains $\mathfrak u(1)^{p-1}$ together with the semisimple algebra obtained by deleting the nodes with $s_i \neq 0$.  This follows since the shift vector satisfies
\bea
\alpha_0\cdot\boldsymbol\delta_L=-1+\frac{s_0}{N}~,\hspace{0.4 in} \alpha_i\cdot\boldsymbol\delta_L=\frac{s_i}{N}~.
\eea
Our tables retain only maximally semisimple left-moving algebras.  Thus, exactly one affine label is nonzero in each simple factor.  This convention excludes branches with an additional left-moving $U(1)$; the universal right-moving Abelian factors of toroidal compactification are a separate matter and are discussed in Section \ref{sec:eightdimensions}.\footnote{Note that right-moving Abelian factors can also be enhanced to non-Abelian ones, as will be discussed below.}

Note that for an automorphism of the Lie algebra of exactly order $N$, the affine labels are chosen to be primitive, i.e. $\mathrm{gcd}(s_0, \dots, s_r) = 1$.\footnote{Indeed, if $\mathrm{gcd}(s_0, \dots, s_r) = d$, then we may write $s_i = d t_i$ and 
\bea
N = \sum_i \kappa_i s_i = d \sum_i \kappa_i t_i = d N'
\eea 
for some integer $N'$. At the same time, the shift vector is given by 
\bea
\boldsymbol{\delta}_L = {1\over N} \sum_{i=1}^r s_i \omega_i = {1\over N'} \sum_{i=1}
^r t_i \omega_i~,
\eea
and hence, recalling that $\alpha_i \cdot \omega_j = \delta_{ij}$, we conclude that the action is of order $N'$.
}  However, in the string construction it is useful to retain certain non-primitive choices. 
In particular, for $N=2$, the equations in (\ref{eq:Kacequations}) admit only two possible maximally-semisimple solutions: \emph{i)} a mark-two node with affine label 1, or \emph{ii)} a mark-one node with affine label $2$. In the first case, we have $s_m=1$ for one $m$ and zero for the others, and hence the labels are primitive. Since $\boldsymbol\delta_L = {\omega_m \over 2} $ and $\alpha_m \cdot \boldsymbol{\delta}_L = \half$, we have 
\bea
E_{\alpha_m} \,\,\mapsto\,\, - E_{\alpha_m}~,
\eea
and hence this is a genuine order-two inner automorphism of the Lie algebra. 

In contrast, in the second case we have $s_m = 2$ for one $m$ and zero for the others, and hence the labels have greatest common divisor two. In this case we have $\boldsymbol\delta_L = {\omega_m } $ and  $e^{2 \pi i \alpha \cdot \omega_m}=1$ for every root, and hence the action on all $E_{\alpha}$ is trivial. Nevertheless, we allow for such shift vectors in our discussion, since they may still give rise to non-trivial actions on other states of the theory via the general formula (\ref{eq:Tdeltadef}).

To summarize, for $N=2$ there are two cases to consider, 
\bea
\boldsymbol\delta_L=\frac{\omega_m}{\kappa_m}~, \hspace{0.4 in} \kappa_m\in\{1,2\}~,
\eea
where $\omega_0$ is the zero vector.  When $\kappa_m=2$, deleting node $m$ gives a proper invariant semisimple subalgebra.  When $\kappa_m=1$, the Kac marks are non-primitive and every root pairs integrally with $\omega_m$, so the adjoint root system is unchanged.  Nevertheless, glue vectors or other lattice weights can still acquire a minus sign.  For a reducible parent $L=\bigoplus_kL_k$, one chooses an allowed node in each factor and sums the mutually orthogonal contributions,
\bea
\label{eq:leftmovingshift}
\boldsymbol\delta_L=\sum_k\frac{\omega^{(k)}_{m_k}}{\kappa^{(k)}_{m_k}}~.
\eea
This gives a finite list of candidate left-moving actions.

As we have just described,   Kac's theorem allows us to obtain the full list of possible left-moving shift vectors for any given parent supersymmetric theory. However, a left-moving shift vector is not yet a string orbifold.  Indeed, we must further require that $\boldsymbol\delta_L$ lift to a vector $\delta=(\boldsymbol\delta_L;\boldsymbol\delta_R)$ obeying (\ref{eq:deltaconstraints}).  To do so, note that since $2\delta$ is a Narain vector, we may write
\bea
\label{eq:delta16wn}
\delta_\ell^I&=&\delta_{16}^I+A_i^I\frac{\hat w^i}{2}~,
\no\\
\delta_{L,a}&=&\frac1{\sqrt2}e^{*i}_a\left[\frac{\hat n_i}{2}+(2G_{ij}-E_{ij})\frac{\hat w^j}{2}-\delta_{16}\cdot A_i\right]~,
\no\\
\delta_{R,a}&=&\frac1{\sqrt2}e^{*i}_a\left[\frac{\hat n_i}{2}-E_{ij}\frac{\hat w^j}{2}-\delta_{16}\cdot A_i\right]~,
\eea
where $2\delta_{16}\in\Gamma_{16}$ and $\hat w^i,\hat n_i\in\mathbb Z$.  Adding a vector of $\Gamma_{16+d,d}$ to $\delta$ does not change its character on the lattice.  Equivalent representatives may therefore shift $\hat w^i$ and $\hat n_i$ by even integers, accompanied by the corresponding change of $\delta_{16}$.  As such, we may effectively regard $\hat{w}^{i}$ and $\hat{n}_{i}$ as $\mathbb{Z}_{2}$-valued variables. 
Note that in terms of these variables, the anomaly-free condition $\delta^2 \in \ZZ$ becomes 
\bea
\label{eq:anomfree2}
    \delta_{16}^2+\half\hat{w}^{i}\hat{n}_{i}\in\mathbb{Z}~.
\eea

Now starting from $\boldsymbol\delta_L$, the lift to $\delta$ is found as follows.  First, for every parity choice of $\hat w^i$, require
\bea
2\delta_{16}=2\delta_\ell-A_i\hat w^i\in\Gamma_{16}~,\hspace{0.4 in} \delta_{16}:=\delta_\ell-\frac12A_i\hat w^i~.
\eea
The second equation of (\ref{eq:delta16wn}) then fixes
\bea
\hat n_i=2\sqrt2e_i^a\delta_{L,a}-(2G_{ij}-E_{ij})\hat w^j+2\delta_{16}\cdot A_i~,
\eea
which must be integral.  Finally, we retain only candidates satisfying (\ref{eq:anomfree2}).
Every surviving tuple $(\delta_{16},\hat w^i,\hat n_i)$ defines a consistent order-two shift orbifold.

\subsection{From moduli to roots and weights}
\label{sec:modulitoroots}
A practical difficulty in the construction of the shift vectors above is that the supersymmetric classification in e.g. \cite{Font:2020rsk} gives the moduli $(G, B, A_i)$ of the compactification, but does not explicitly give the root basis. As such, this must be recovered manually by solving the gauge enhancement conditions in (\ref{eq:gaugeenhancement}). More explicitly, we must carry out the following steps,

\begin{enumerate}
\item Solve $P_L^2=2$ and $P_R=0$.  Using (\ref{eq:Pinnerprod}), these conditions become
\bea
\Pi^2+2w^in_i=2~, \hspace{0.4 in} n_i=E_{ij}w^j+\Pi\cdot A_i~.
\eea
The resulting vectors form the root system of a rank-$(16+d)$ ADE algebra
\bea
\Ls=\bigoplus_{k=1}^nL_k~, \hspace{0.4 in}\sum_{k=1}^nr_k=16+d~.
\eea
\item Choose a positive system and extract simple roots
\bea
\Sigma=\bigsqcup_{k=1}^n\Sigma_k~,\hspace{0.4 in}\Sigma_k=\{\alpha^{(k)}_1,\ldots,\alpha^{(k)}_{r_k}\}~.
\eea
\item In each factor determine the lowest root
\bea
\alpha^{(k)}_0=-\sum_{i=1}^{r_k}\kappa^{(k)}_i\alpha^{(k)}_i~,\hspace{0.4 in} \kappa^{(k)}_i\in\mathbb Z_{>0}~,
\eea
and hence the Kac marks.
\item Determine the fundamental weights from
\bea
\omega^{(k)}_i\cdot\alpha^{(\ell)}_j=\delta_{k\ell}\delta_{ij}~.
\eea
\end{enumerate}
These weights can then be inserted into (\ref{eq:leftmovingshift}), after which the Narain-lattice lift and level-matching tests described above are applied.

\subsection{Performing the $\mathbb Z_2$ orbifold}
\label{sec:orbifold}
For every maximally enhanced supersymmetric starting point, the preceding steps produce a finite set of anomaly-free symmetries $g_\delta=(-1)^\F\T_\delta$.  From (\ref{eq:Pinnerprod}), their action on a state labelled by $(\Pi,w^i,n_i)$ is given in terms of
\bea
\label{eq:shift-character}
P\cdot\delta=\Pi\cdot\delta_{16}+\frac12\bigl(n_i\hat w^i+w^i\hat n_i\bigr)~.
\eea
Consequently, $\hat w^i=1$ produces the phase $(-1)^{n_i}$ associated with a geometric half-period translation along the $i$-th circle, whereas $\hat n_i=1$ produces $(-1)^{w^i}$ and is a half-translation in the T-dual coordinate.  The vector $\delta_{16}$ is a gauge-lattice translation.  A general shift combines these operations and can be asymmetric or non-geometric in a given polarization \cite{Narain:1986qm}.\footnote{{The variables $\hat w^i$ and $\hat n_i$ determine whether supersymmetry is restored in the decompactification limit and play a role in constructing interpolating models that can exhibit an exponentially suppressed cosmological constant \cite{Itoyama:1986ei,Itoyama:1987rc,Itoyama:2021itj,Koga:2022qch}.}
}

Let $h\in\{0,1\}$ label the spatial twist and $g\in\{0,1\}$ the temporal twist.  The Narain-lattice block is
\bea
\label{eq:bosonicblock}
\Lambda^g{}_h(\delta;\tau):= \sum_{P\in\Gamma_{16+d,d}+h\delta} e^{2\pi ig\left(P\cdot\delta-\frac h2\delta^2\right)} q^{P_L^2/2}\bar q^{P_R^2/2}~, \hspace{0.4 in} q=e^{2\pi i\tau}~.
\eea
This is obtained as follows. First, the case of $g=h=0$ is the ordinary Narain theta function.  Inserting $\T_\delta$ then gives a factor of $e^{2\pi iP\cdot\delta}$ in the untwisted trace.  Next, a modular $S$ transformation, or equivalently a Poisson resummation together with self-duality of the lattice, maps this block to a sum over the shifted coset $\Gamma_{16+d,d}+\delta$.  Finally, another modular $T$ transformation multiplies the states by $e^{\pi iP^2}$, and writing $P=Q+\delta$ with $Q\in\Gamma_{16+d,d}$ and using $Q^2\in2\mathbb Z$ gives
\bea
 e^{\pi iP^2} =e^{\pi i(Q^2+2Q\cdot\delta+\delta^2)} =e^{2\pi i(Q\cdot\delta+\delta^2/2)}  =e^{2\pi i(P\cdot\delta-\delta^2/2)}~,
\eea
which fixes the cocycle in (\ref{eq:bosonicblock}).  The condition $\delta^2\in\mathbb Z$ required in (\ref{eq:deltaconstraints}) makes the twisted insertion square consistently and closes the four blocks under modular transformations \cite{Dixon:1985jw,Narain:1986qm}.

On the other hand, the GSO-projected fermion blocks take the following form, 
\bea
\mathcal R^g{}_h(\bar\tau):= {1 \over 2\bar\eta^4} \sum_{s_1,s_2=0}^1 (-1)^{(1+g)s_1+(1+h)s_2+s_1s_2+gh} \, \bar\vartheta^4\!\left[\begin{matrix}s_1/2\\s_2/2\end{matrix}\right](\bar\tau)~.
\eea
To see this, we begin with the case of $g=h=0$, which is the GSO-projected block with standard choice of signs. Since $\mathrm{NS}$ sector states are bosons while $\mathrm{R}$ sector states are fermions, inserting $(-1)^\F$  corresponds to insertion of a factor of $(-1)^{s_1}$ in the sum. A modular $S$ transformation then exchanges the roles of $s_1$ and $s_2$, and the full twisted-twined block is finally obtained by a modular $T$-transformation.

Including the non-compact bosons and oscillators, the one-loop partition function of the orbifolded theory is then given by
\bea
\label{eq:nonSUSYpartfunc}
Z_{\cancel{\mathrm{SUSY}}}^{(\delta)}(\tau) ={1 \over 2\tau_2^{(8-d)/2}\eta^{24}\bar\eta^8}
\sum_{g,h=0}^1\mathcal R^g{}_h(\bar\tau)\Lambda^g{}_h(\delta;\tau)~.
\eea
This partition function may be rewritten in a somewhat more convenient form as follows. We begin by reorganizing the fermionic blocks into level-one $SO(8)$ characters via
\bea
\mathcal R^0{}_0&=&\bar V_8-\bar S_8~, \hspace{0.55 in} \mathcal R^1{}_0\,\,\,\,=\,\,\,\,\bar V_8+\bar S_8~,
\no\\
\mathcal R^0{}_1&=&\bar O_8-\bar C_8~, \hspace{0.5 in} \mathcal R^1{}_1\,\,\,\,=\,\,\,\,-\bigl(\bar O_8+\bar C_8\bigr)~.
\eea
Then defining
\bea
\Gamma_{16+d,d}^+(\delta)&:=&\{Q\in\Gamma_{16+d,d}\mid Q\cdot\delta\in\mathbb Z\}~,
\no\\
\Gamma_{16+d,d}^-(\delta)&:=&\{Q\in\Gamma_{16+d,d}\mid Q\cdot\delta\in\mathbb Z+\tfrac12\}~,
\eea
it is straightforward to check that 
\begin{align}
\label{eq:projected-partition}
Z_{\cancel{\mathrm{SUSY}}}^{(\delta)}(\tau)
=\frac1{\tau_2^{(8-d)/2}\eta^{24}\bar\eta^8}\Bigg[
&\bar V_8\!\sum_{Q\in\Gamma^+(\delta)}q^{Q_L^2/2}\bar q^{Q_R^2/2}
-\bar S_8\!\sum_{Q\in\Gamma^-(\delta)}q^{Q_L^2/2}\bar q^{Q_R^2/2}\nonumber\\
&+\bar O_8\!\sum_{P\in\Gamma_O(\delta)+\delta}q^{P_L^2/2}\bar q^{P_R^2/2}
-\bar C_8\!\sum_{P\in\Gamma_C(\delta)+\delta}q^{P_L^2/2}\bar q^{P_R^2/2}
\Bigg]~,
\end{align}
where we have defined 
\bea
\Gamma_O(\delta)&:=&
\begin{cases}
\Gamma^+(\delta),&\delta^2\text{ odd}\\
\Gamma^-(\delta),&\delta^2\text{ even}
\end{cases}
\qquad
\Gamma_C(\delta):=
\begin{cases}
\Gamma^-(\delta),&\delta^2\text{ odd}\\
\Gamma^+(\delta),&\delta^2\text{ even}
\end{cases}~.
\eea

\subsection{Tachyons and massless states}
\label{sec:light-states}
Finally, it is interesting to ask whether the non-supersymmetric theories constructed above have tachyons or massless states.
Massless spacetime fermions arise from the $\bar S_8$ and $\bar C_8$ Ramond ground states.  In units $\alpha'=1$, their lattice momenta satisfy
\bea
\label{eq:masslessfermioncond}
P_R=0~,\hspace{0.4 in}P_L^2=2~,
\eea
with $P\in\Gamma^-(\delta)$ in the  $S_8$ sector and $P\in\Gamma_C(\delta)+\delta$ in the $C_8$ sector.  The number $N_0^{\rm f}$ is the total multiplicity from these sectors; the multiplicity carried by the $SO(8)$ character is not included.

Additional massless right-moving vectors may arise from the $\bar O_8$ states.  The conditions are given by
\bea
\label{eq:rightmovingmasslessvectorcond}
P_R^2=1~,\hspace{0.4 in}P_L=0~.
\eea
Note that the only possible non-Abelian factors arising from the enhancement are of type $A_1$.

Tachyonic states arise from the $\bar O_8$ ground state in the shifted sector.  A vector $P\in\Gamma_O(\delta)+\delta$ gives a tachyonic or massless scalar precisely when
\bea
\label{eq:tachyonmasslessscalarcond}
0\leq P_R^2\leq1~, \hspace{0.4 in} P_L^2=P_R^2+1~,\hspace{0.4 in} m^2=2(P_R^2-1)~.
\eea
Thus $P_R^2<1$ is tachyonic, while $P_R^2=1$ is a massless scalar at the boundary of a tachyonic region.  The quantity $N_0^{\rm s}$ below counts these shifted-sector $O_8$ lattice states.  It does not include the universal dilaton or neutral toroidal and Cartan moduli.

In general, to identify all tachyons and massless scalars, we must scan over all $P= (P_L, P_R)$, or equivalently all $(w^i, n_i, \Pi)$ in the notation of (\ref{eq:ellppdef}), and identify those that satisfy (\ref{eq:tachyonmasslessscalarcond}). To make this scan more amenable to computer implementation, it is useful to obtain bounds on the data $(w^i, n_i, \Pi)$, which combined with the fact that $w^i,n_i,$ and $\Pi$ are vectors of integers makes the scan finite. 

To obtain such bounds, let us define
\bea
\label{eq:shiftedvariables}
\widetilde w^i:=w^i+\frac{\hat w^i}{2}~,\hspace{0.4 in} \widetilde n_i:=n_i+\frac{\hat n_i}{2}~, \hspace{0.4 in} \widetilde\Pi:=\Pi+\delta_{16}~.
\eea
In terms of these variables, the level-matching equation becomes
\bea
\label{eq:PLPRcondition}
1=P_L^2-P_R^2=2\widetilde n_i\widetilde w^i+\widetilde\Pi^2~.
\eea
Next we define
\bea
\eta_i:=\widetilde n_i-E_{ij}\widetilde w^j-\widetilde\Pi\cdot A_i~,
\qquad
P_R^2=\frac12\eta_i(G^{-1})^{ij}\eta_j~.
\eea
Substituting for $\widetilde n_i$ in (\ref{eq:PLPRcondition}) and using the antisymmetry of $B$ gives
\bea
\label{eq:bigwGwconstraint}
1=2\widetilde w^iG_{ij}\widetilde w^j
+\bigl(\widetilde\Pi+A_i\widetilde w^i\bigr)^2
+2\eta_i\widetilde w^i~.
\eea
Because $P_R^2\leq1$, one has $\eta^TG^{-1}\eta\leq2$. The Cauchy-Schwarz inequality applied to the last term of (\ref{eq:bigwGwconstraint}) then yields two bounds
\bea
\label{eq:finite-bounds}
\widetilde w^TG\widetilde w
&\leq&\left(1+\frac1{\sqrt2}\right)^2=\frac32+\sqrt2~,
\no\\
\left\|\widetilde\Pi+A_i\widetilde w^i\right\|^2
&\leq&1-2\widetilde w^TG\widetilde w
+2\sqrt2\sqrt{\widetilde w^TG\widetilde w}~.
\eea
The first inequality confines $w\in\mathbb Z^d$ to a finite ellipsoid.  For each such $w$, the second confines $\Pi$ to finitely many vectors in the positive-definite gauge lattice, while (\ref{eq:PLPRcondition}) and the right-moving bound leave only finitely many $n_i$.

We thus have the following finite search to perform in arbitrary dimension $d$,
\begin{enumerate}
\item Enumerate all $w\in\mathbb Z^d$ satisfying
\bea
\left(w+\frac{\hat w}{2}\right)^TG\left(w+\frac{\hat w}{2}\right)\leq\frac32+\sqrt2~.
\eea
\item For each $w$, enumerate all $\Pi\in\Gamma_{16}$ satisfying
\bea
\left(\Pi+\delta_{16}+A_i\left(w^i+\frac{\hat w^i}{2}\right)\right)^2
\leq F\left(w+\frac{\hat w}{2}\right)~,
\eea
where
\bea
F(x):=1-2x^TGx+2\sqrt2\sqrt{x^TGx}~.
\eea
\item Solve (\ref{eq:PLPRcondition}) for the remaining integer momenta, impose $0\leq P_R^2\leq1$, and retain only vectors in $\Gamma_O(\delta)+\delta$.
\item Classify the retained states as tachyonic for $P_R^2<1$ and massless for $P_R^2=1$.  Determine massless fermions independently from (\ref{eq:masslessfermioncond}) and massless right-moving vectors from (\ref{eq:rightmovingmasslessvectorcond}).
\end{enumerate}

For a tachyon-free theory we evaluate the one-loop cosmological constant
\bea
\label{eq:cosmological-constant}
\Lambda^{(10-d)}(\delta):=-\frac12(4\pi^2\alpha')^{-(10-d)/2}
\int_{\mathcal F}\frac{d^2\tau}{\tau_2^2}\,
Z_{\cancel{\mathrm{SUSY}}}^{(\delta)}(\tau)~,
\eea
where $\mathcal F$ is the modular fundamental domain.  Note that the sign of $\Lambda$ alone is not a stability criterion.  A massless $O_8$ scalar is a knife-edge state in the terminology of \cite{Ginsparg:1986wr}---an infinitesimal deformation can enter an adjacent tachyonic region.

For a scalar-free entry, the next local diagnostic is the Hessian, which at a maximally enhanced point takes the form
\bea
\label{eq:hessian}
{(H^{(\delta)})^{a}}_b:=g^{a c} \frac{\partial^2}{\partial\phi_c\partial\phi_b}\Lambda^{(10-d)}(\delta)~,
\eea
where $\phi_a$ denote the $d^2$ massless moduli of the torus as well as the $16d$ Wilson line moduli, and $g^{ab}$ is the inverse metric on moduli space; see Appendix \ref{app:modulispace} for additional details. 
By computing the eigenvalues of the Hessian numerically, we are able to diagnose whether a given point is a local minimum, a local maximum, or a saddle point. 

\section{Results}

In this section, we use the machinery described in the previous section to classify eight- and nine-dimensional maximally enhanced non-supersymmetric string theories. For the latter, we confirm that the results match with the known ones in \cite{Fraiman:2023cpa}. The ten-dimensional case is also reviewed in Appendix \ref{app:10d}. 

\subsection{Nine dimensions}
\label{sec:nine-dimensions}

We begin in nine dimensions, for which the maximally enhanced supersymmetric theories were classified in \cite{Font:2020rsk}. These theories admit a convenient description as follows.  Choosing an $E_8\times E_8$ polarization of the even self-dual lattice $\Gamma_{17,1}$, the maximally enhanced supersymmetric circle points can be obtained by deleting one node from each of the two affine $E_8$ diagrams while retaining the central circle node.\footnote{Using the $(E_8\times E_8)\rtimes \ZZ_2$ frame is only a choice of polarization; the resulting Narain points may equally well be written in the $\operatorname{Spin}(32)/\mathbb Z_2$ frame.
}  Let $\omega_k$ and $\kappa_k$ be the fundamental weight and Kac mark associated with node $k$, with $\omega_0=0$ and $\kappa_0=1$.  A convenient representative of the point labelled by $(k,k')$ is
\bea
\label{eq:9dparents}
   A(k,k')=
   \frac{\omega_k}{\kappa_k}\oplus \frac{\omega_{k'}}{\kappa_{k'}}~, \hspace{0.5 in}  E=R^2+\frac12A^2=1~. 
\eea
Taking $0\leq k\leq k'\leq8$, identifying the exchange of the two $E_8$ factors, and excluding $(k,k')=(8,8)$ for which $R=0$, we are left with $44$ maximally enhanced supersymmetric parent points.

For each parent point, we reconstruct the simple roots and fundamental weights from the moduli $(R,A)$ by the procedure of Section \ref{sec:modulitoroots}.  We then form all candidate vectors $\boldsymbol\delta_L$ in (\ref{eq:leftmovingshift}) and test whether each candidate admits a full Narain-lattice lift $\delta$ satisfying (\ref{eq:deltaconstraints}).

As a concrete example, take the parent theory $(k,k') = (0,0)$, which has 
\bea
A=0~, \hspace{0.3 in} R=1~, \hspace{0.3 in}  L_{\rm SUSY}=A_1+E_8+E_8~.
\eea
Choose the trivial action on the $A_1$ factor and place the nonzero Kac label on node $8$ in each $E_8$.  Since $\kappa_8=2$ (see Appendix \ref{app:10d}), the candidate vector is
\bea
\label{eq:candidatedeltaL}
\boldsymbol{\delta}_L =0_{A_1}\oplus\frac{\omega_8}{2} \oplus\frac{\omega'_8}{2}~. 
\eea
Comparing with (\ref{eq:delta16wn}), one consistent lift is
\bea
\hat w = \hat n = 0~, \hspace{0.3 in} \delta_{16} = \delta_\ell~, 
\eea
for which $\boldsymbol{\delta}_R=0$.  In the conventions of Appendix \ref{app:10d}, $\omega_8^2=(\omega'_8)^2=4$, so we have 
\bea
 \delta^2=\frac14\omega_8^2
   +\frac14(\omega'_8)^2=2\in\mathbb Z~.
\eea
This is integral and therefore satisfies the remaining condition in (\ref{eq:deltaconstraints}).  The candidate $\boldsymbol\delta_L$ in (\ref{eq:candidatedeltaL}) consequently lifts to a consistent Narain shift and defines a maximally enhanced non-supersymmetric theory.

The invariant algebra follows directly from Section \ref{sec:Kactheorem}.  In the standard coordinate realization, $\omega_8/2=(0^7,1)$.  This vector pairs integrally with the $112$ roots $\pm e_i\pm e_j$ of $D_8$ and half-integrally with the $128$ spinorial roots of $E_8$.  The spinorial roots are therefore projected out, while the $D_8$ roots survive.  The two circle roots also have trivial phase and supply the untouched $A_1$.  Thus (\ref{eq:DeltaLndef}) gives
\bea
   \Ln=A_1+2D_8~, 
\eea
with $2+112+112=226$ invariant roots.  This matches with entry $41$ of the classification in \cite{Fraiman:2023cpa}, and is one of its tachyon-free saddle points.

Likewise, choosing node $8$ in only one $E_8$ factor instead gives $A_1+D_8+E_8$, which matches with entry $103$ in \cite{Fraiman:2023cpa}.  
The complete scan over the $44$ parents and all consistent lifts reproduces the $95$ maximally semisimple entries of \cite{Fraiman:2023cpa}.  The qualifier ``semisimple'' is essential: the direct classification contains twelve further reflection-fixed points with an additional left-moving $U(1)$ factor, which are excluded by our one-node-per-simple-factor convention.  Table \ref{tab:9dmatching} records representative entries of the dictionary; the full classification can be found in Appendix \ref{app:8d}. The pair $(k,k')$ labels the supersymmetric parent through (\ref{eq:9dparents}), and the last column gives the entry number in \cite{Fraiman:2023cpa}.

\begin{table}[t]
    \centering
    \small
    \begin{tabular}{|c|c|c|c|} \hline
         $(k,k')$ & $\Ls$ & $\Ln$ & \cite{Fraiman:2023cpa} entry
         \\ \hline\hline
       \multirow{3}{*}{$(0,0)$} & \multirow{3}{*}{$A_1+2E_8$} & $A_1+2D_8$ & $41$ \\
           &  & $3A_1+2E_7$ & $97$ \\
           &  &  $A_1+D_8+E_8$ & $103$ \\ \hline
       $(0,1)$ & $A_9+E_8$ & $A_9+D_8$ & $46$ \\ \hline
         \multirow{2}{*}{$(0,2)$} & \multirow{2}{*}{$A_1+A_2+A_6+E_8$} &  $A_1+A_2+A_6+D_8$ & $44$ \\ 
         &  &  $2A_1+A_2+A_6+E_7$ & $84$ \\ \hline
        $(0,3)$ & $A_4+A_5+E_8$  & $A_4+A_5+D_8$ & $43$ \\ \hline
        \multirow{4}{*}{$(0,4)$} & \multirow{4}{*}{$A_4+D_5+E_8$}
        & $A_4+D_5+D_8$ & $36$ \\ 
           &  &  $A_4+D_5+D_8$ & $47$ \\ 
           &  &  $3A_1+A_3+A_4+E_7$ & $80$ \\ 
           &  &  $A_4+D_5+E_8$ & $101$ \\ \hline
        \multirow{2}{*}{$(0,5)$} & \multirow{2}{*}{$A_3+D_6+E_8$} &  $A_3+D_8+E_6$ & $78$ \\ 
          & & $2A_1+A_3+A_5+E_7$ & $82$ \\ \hline
    \end{tabular}
    \caption{Representative entries in the nine-dimensional matching test.  The parent label $(k,k')$ fixes the supersymmetric Narain point.  The final column gives the corresponding entry in the direct extended-Dynkin classification of \cite{Fraiman:2023cpa}. For simplicity, this table only contains cases with $\hat{w}=\hat{n}=0$ mod 2; accounting for non-zero cases gives additional possibilities such as $\Ln=2A_1+E_7+E_8$ or $2A_1+E_7+D_8$ for $(k,k') = (0,0)$. The full list can be found in Appendix \ref{app:8d}.  }
    \label{tab:9dmatching}
\end{table}

The agreement is stronger than a match of abstract gauge algebras: the shift construction also reproduces the spectrum of light states such as tachyons and massless scalars, as well as the correct cosmological constant in the tachyon-free cases.  For example, entries $36$ and $47$ of \cite{Fraiman:2023cpa} both have algebra $A_4+D_5+D_8$, but they describe distinct backgrounds: entry $36$ is tachyon-free, whereas entry $47$ contains tachyons. The current construction successfully reproduces this distinction. Furthermore, note that the same $\Ln$ can arise from different $L_{\rm SUSY}$. For instance, $\Ln=A_1+2D_8$ can be realized starting from both $\Ls=A_1+2E_8$ and $A_1+D_{16}$. Throughout the remainder of this paper, we retain only physically distinct configurations and do not count multiple realizations of the same background separately.

\subsection{Eight dimensions}
\label{sec:eightdimensions}

We next proceed to the case of eight dimensions, where the set of supersymmetric maximally-enhanced theories was identified in \cite{Font:2020rsk}. In this case the corresponding set of non-supersymmetric theories has not yet appeared in the literature.

 For each parent we use the known moduli $(G,B,A_i)$ to reconstruct its rank-eighteen ADE root system as in Section \ref{sec:modulitoroots}.  We then choose one affine node in each simple factor, form the candidate $\boldsymbol\delta_L$ of (\ref{eq:leftmovingshift}), and solve (\ref{eq:delta16wn}) for all compatible tuples $(\delta_{16},\hat w^i,\hat n_i)$.  Candidates that fail $2\delta\in\Gamma_{18,2}$ or $\delta^2\in\mathbb Z$ are discarded.  For every surviving shift, the invariant roots in (\ref{eq:DeltaLndef}) determine the non-supersymmetric left-moving algebra, and the finite search of Section \ref{sec:light-states} determines whether the model contains a tachyon or massless scalars.

For each consistent lift we compute four pieces of data.  First, we enumerate the invariant norm-two vectors and identify their root system.  Second, we perform the shifted-sector search of Section \ref{sec:light-states} to identify tachyonic states.  Third, we count the massless lattice vectors in the $S_8$ and $C_8$ sectors, which give the number of massless fermions.  Fourth, for each tachyon-free model we evaluate the modular integral in (\ref{eq:cosmological-constant}) using the full partition function. 
Finally, for models which are free of both tachyons and massless scalars, we compute the Hessian and use this to compute the ratio ${\left(g^{ac}\nabla_c\nabla_b V\right)/{V}}$ relevant for the refined de Sitter swampland conjecture.

While conceptually straightforward, this algorithm turns out to be computationally intensive. In particular, although the orbifold partition function is given in closed form as in \eqref{eq:projected-partition}, obtaining the $q$-expansion required for the computation of the cosmological constant, as well as the Hessian in the absence of massless scalars, is challenging.
Even when restricting the expansion to $O(q^3 \bar{q}^3)$, enumerating all elements in the $20$-dimensional Narain lattice $\Gamma_{18,2}$ that contribute to this order remains highly non-trivial. In practice, the computation took $O(10^5)$ CPU hours in total, which was completed within $3$ weeks by utilizing the Genkai cluster at Kyushu University. Note that in total there are 
 1210 cases, but we only give the tachyon-free ones below; data for the remaining cases is included as a separate ancillary file. 

Table \ref{tab:8d} contains the $71$ entries with no tachyon.  The table is deliberately condensed; Appendix \ref{app:8d} gives a full representative of the shift and the supersymmetric parent algebra for every row.  The columns should be read as follows.
\begin{itemize}[leftmargin=1.8em]
\item $\#$ is a cross-reference shared by  Table \ref{tab:8d} and Appendix \ref{app:8d}.  It has no  physical meaning.
\item $k$ labels the parent supersymmetric Narain point, and matches with the label given in Table 12 of \cite{Font:2020rsk}.  Note that for entries in which the value of $k$ is accompanied by a number in parentheses, the corresponding $L_{\mathrm{SUSY}}$ in Table 12 of \cite{Font:2020rsk} is realized by more than one choice of moduli. The number in parentheses specifies which of these moduli configurations, counted from top to bottom, is used for that entry.

\item $\Ln$ denotes the semisimple left-moving current algebra generated by invariant norm-two vectors.  It has rank $18$ in this scan.  The full eight-dimensional gauge algebra also contains the two universal right-moving factors $\mathfrak u(1)^2_R$, which are common to most rows and therefore omitted (note, however, that these factors may be enhanced to a non-Abelian symmetry, in which case we write it as $A_1^{(\mathrm R)}$).  Accordingly, ``no Abelian factor'' means no additional \emph{left-moving} $\mathfrak u(1)$.  
Shifts obtained by deleting more than one affine node in a simple parent factor retain such a left-moving Abelian current and lie outside our maximally semisimple data set.

\item $N_0^{\rm s}$ is the lattice multiplicity of shifted-sector $O_8$ states with $P_R^2=1$.  Thus $N_0^{\rm s}>0$ marks a knife-edge point, whereas $N_0^{\rm s}=0$ only removes this particular tree-level instability and does not establish one-loop stability.
\item $N_0^{\rm f}$ is the total lattice multiplicity of massless Ramond-sector states in the $S_8$ and $C_8$ sectors, described below (\ref{eq:masslessfermioncond}).  The multiplicity of the $SO(8)$ characters is not included.
\item $\Lambda^{(8)}$ is the numerical value of (\ref{eq:cosmological-constant}) for $d=2$, evaluated with $\alpha'=1$.   Values are shown to two decimal places. 
\end{itemize}
Several rows have the same abstract algebra without describing the same string background.  Distinct embeddings of the shift in $\Gamma_{18,2}$ can produce different sector assignments, massless spectra, and vacuum energies.  It is also interesting to note that some cases have identical cosmological constants, and even identical partition functions, though the organization of states when split up into sectors remains distinct.

Of the $71$ tachyon-free entries, $25$ have $N_0^{\rm s}=0$ and appear separately in Table \ref{tab:intro8d}; the remaining $46$ are knife-edge points.  In our normalization, $68$ entries have positive $\Lambda^{(8)}$ and three have negative $\Lambda^{(8)}$.  Furthermore, explicit computation of the Hessian reveals that one case (case No. 2) is a local maximum, while all the others are saddle points. 
For example, in case No. 12, which has $\Ln = A_1 + A_7 + 2D_5$, the Hessian eigenvalues $\times 10^{6}$ and their multiplicities (written as exponents) are found to be 
\bea
&\vphantom{.}& 296.92^{1}, -101.55^{17}, -98.775^{17}, 
-41.852^{1}~. \quad 
\eea
This is almost a local maximum, but the existence of a single positive eigenvalue makes it into a saddle point.  The eigenvalues of the Hessian for all scalar-free entries are listed in Table \ref{tab:hessian8d}.

\begingroup
\scriptsize
\begin{longtable}{@{}c c p{0.29\textwidth} c c c@{}}
\caption{Condensed summary of the 71 tachyon-free eight-dimensional entries. The complete Narain-shift representatives and parent algebras are given in Appendix \ref{app:8d}. }\label{tab:8d}\\
\toprule
$\#$ & $k$ & $\Ln$ & $N_0^{\rm s}$ & $N_0^{\rm f}$ & $\Lambda^{(8)}\times 10^6$ \\
\midrule
\endfirsthead
\toprule
$\#$ & $k$ & $\Ln$ & $N_0^{\rm s}$ & $N_0^{\rm f}$ & $\Lambda^{(8)}\times 10^6$ \\
\midrule
\endhead
\midrule
\multicolumn{6}{r}{\emph{continued on next page}}\\
\endfoot
\bottomrule
\endlastfoot
1 & $1$ & $\scriptstyle 6A_{3}+A_1^{(\rm R)}$ & 360 & 0 & 63.20 \\
2 & $1$ & $\scriptstyle 6A_{3}$ & 0 & 108 & 106.82 \\
3 & $13$ & $\scriptstyle 3A_{6}+A_1^{(\rm R)}$ & 252 & 0 & 67.41 \\
4 & $25$ & $\scriptstyle 4A_{1} + 2A_{7}$ & 0 & 156 & 106.82 \\
5 & $28$ & $\scriptstyle 2A_{1} + 3A_{3} + A_{7}$ & 0 & 130 & 106.30 \\
6 & $54(1)$ & $\scriptstyle 2A_{9}+2A_1^{(\rm R)}$ & 720 & 0 & 19.59 \\
7 & $83(1)$ & $\scriptstyle 2A_{2} + A_{3} + A_{11}+A_1^{(\rm R)}$ & 312 & 0 & 66.09 \\
8 & $88$ & $\scriptstyle A_{2} + A_{5} + A_{11}+A_1^{(\rm R)}$ & 336 & 0 & 71.57 \\
9 & $89$ & $\scriptstyle A_{1} + A_{6} + A_{11}+A_1^{(\rm R)}$ & 352 & 0 & 61.36 \\
10 & $111(2)$ & $\scriptstyle A_{1} + A_{17}+2A_1^{(\rm R)}$ & 1232 & 0 & $-$87.21 \\
11 & $113$ & $\scriptstyle 2A_{4} + 2D_{5}$ & 0 & 100 & 145.97 \\
12 & $121$ & $\scriptstyle A_{1} + A_{7} + 2D_{5}$ & 0 & 170 & 147.45 \\
13 & $122$ & $\scriptstyle A_{1} + A_{2} + A_{3} + A_{7} + D_{5}$ & 0 & 130 & 130.69 \\
14 & $123$ & $\scriptstyle 2A_{1} + A_{4} + A_{7} + D_{5}$ & 0 & 110 & 120.48 \\
15 & $124$ & $\scriptstyle A_{8} + 2D_{5}$ & 0 & 100 & 111.94 \\
16 & $136$ & $\scriptstyle 3D_{6}+2A_1^{(\rm R)}$ & 720 & 0 & 19.59 \\
17 & $137$ & $\scriptstyle 2A_{1} + 2A_{3} + D_{4} + D_{6}$ & 448 & 128 & 90.05 \\
18 & $137$ & $\scriptstyle 2A_{3} + 2D_{6}$ & 0 & 180 & 106.82 \\
19 & $137$ & $\scriptstyle 4A_{1} + 2A_{3} + 2D_{4}$ & 256 & 176 & 105.88 \\
20 & $154$ & $\scriptstyle A_{7} + D_{5} + D_{6}$ & 0 & 190 & 106.30 \\
21 & $167$ & $\scriptstyle A_{1} + A_{3} + A_{5} + D_{4} + D_{5}$ & 416 & 128 & 96.83 \\
22 & $169$ & $\scriptstyle 2A_{1} + 2D_{8}$ & 1024 & 256 & $-$1.87 \\
23 & $169$ & $\scriptstyle 2A_{1} + 4D_{4}$ & 512 & 256 & 104.94 \\
24 & $177$ & $\scriptstyle 2D_{5} + D_{8}$ & 0 & 228 & 106.82 \\
25 & $177$ & $\scriptstyle 2D_{5} + D_{8}$ & 640 & 128 & 74.22 \\
26 & $177$ & $\scriptstyle 4A_{1} + A_{3} + D_{5} + D_{6}$ & 352 & 176 & 97.96 \\
27 & $177$ & $\scriptstyle 2D_{4} + 2D_{5}$ & 256 & 224 & 105.88 \\
28 & $179$ & $\scriptstyle 2D_{9}$ & 0 & 324 & 106.82 \\
29 & $179$ & $\scriptstyle D_{4} + D_{5} + D_{9}$ & 448 & 224 & 90.05 \\
30 & $187$ & $\scriptstyle A_{4} + D_{5} + D_{9}$ & 0 & 180 & 117.13 \\
31 & $187$ & $\scriptstyle A_{4} + D_{4} + 2D_{5}$ & 160 & 160 & 124.00 \\
32 & $188$ & $\scriptstyle 2A_{1} + 2A_{3} + D_{4} + D_{6}$ & 256 & 208 & 105.88 \\
33 & $188$ & $\scriptstyle 4A_{1} + 2A_{3} + D_{8}$ & 256 & 240 & 105.88 \\
34 & $299$ & $\scriptstyle 2A_{1} + 2A_{4} + D_{8}$ & 0 & 192 & 142.49 \\
35 & $191$ & $\scriptstyle 3A_{1} + A_{5} + D_{4} + D_{6}$ & 576 & 192 & 79.66 \\
36 & $191$ & $\scriptstyle 5A_{1} + A_{5} + D_{8}$ & 384 & 248 & 98.28 \\
37 & $195$ & $\scriptstyle A_{1} + A_{2} + D_{4} + D_{5} + D_{6}$ & 192 & 240 & 131.32 \\
38 & $195$ & $\scriptstyle 3A_{1} + A_{2} + A_{3} + D_{4} + D_{6}$ & 192 & 216 & 131.32 \\
39 & $195$ & $\scriptstyle 3A_{1} + A_{2} + D_{5} + D_{8}$ & 128 & 232 & 129.16 \\
40 & $199$ & $\scriptstyle A_{1} + A_{2} + A_{4} + D_{5} + D_{6}$ & 0 & 184 & 155.53 \\
41 & $199$ & $\scriptstyle A_{1} + A_{2} + A_{3} + A_{4} + D_{8}$ & 0 & 224 & 154.51 \\
42 & $256$ & $\scriptstyle 2A_{1} + 2A_{2} + 2D_{6}$ & 0 & 272 & 166.26 \\
43 & $316$ & $\scriptstyle 2A_{1} + 2A_{2} + D_{4} + D_{8}$ & 0 & 288 & 166.26 \\
44 & $203$ & $\scriptstyle A_{1} + A_{2} + A_{3} + D_{4} + D_{8}$ & 256 & 304 & 133.48 \\
45 & $203$ & $\scriptstyle A_{1} + A_{2} + A_{3} + 2D_{6}$ & 128 & 208 & 129.16 \\
46 & $204$ & $\scriptstyle 2A_{1} + A_{4} + D_{4} + D_{8}$ & 256 & 288 & 124.78 \\
47 & $204$ & $\scriptstyle 2A_{1} + A_{4} + 2D_{6}$ & 256 & 272 & 124.78 \\
48 & $205$ & $\scriptstyle A_{1} + D_{4} + D_{5} + D_{8}$ & 256 & 336 & 148.43 \\
49 & $205$ & $\scriptstyle 3A_{1} + A_{3} + 2D_{6}$ & 256 & 296 & 148.43 \\
50 & $206$ & $\scriptstyle D_{4} + D_{6} + D_{8}$ & 512 & 352 & 104.94 \\
51 & $206$ & $\scriptstyle 2A_{1} + D_{4} + 2D_{6}$ & 512 & 304 & 104.94 \\
52 & $314$ & $\scriptstyle A_{1} + A_{4} + D_{5} + D_{8}$ & 0 & 288 & 175.58 \\
53 & $315$ & $\scriptstyle A_{5} + D_{5} + D_{8}$ & 0 & 288 & 147.19 \\
54 & $209$ & $\scriptstyle 2A_{1} + A_{3} + D_{6} + D_{7}$ & 256 & 256 & 105.88 \\
55 & $209$ & $\scriptstyle 2D_{5} + D_{8}$ & 256 & 288 & 105.88 \\
56 & $316$ & $\scriptstyle 2A_{2} + D_{6} + D_{8}$ & 0 & 320 & 166.26 \\
57 & $211$ & $\scriptstyle 2A_{1} + A_{2} + D_{6} + D_{8}$ & 0 & 384 & 187.77 \\
58 & $212$ & $\scriptstyle A_{1} + A_{3} + D_{6} + D_{8}$ & 0 & 384 & 165.36 \\
59 & $213$ & $\scriptstyle A_{4} + D_{6} + D_{8}$ & 256 & 320 & 124.78 \\
60 & $214$ & $\scriptstyle A_{1} + A_{2} + D_{7} + D_{8}$ & 256 & 352 & 133.48 \\
61 & $214$ & $\scriptstyle A_{1} + A_{2} + D_{6} + D_{9}$ & 128 & 280 & 129.16 \\
62 & $296$ & $\scriptstyle 2A_{1} + 2D_{8}$ & 0 & 512 & 211.74 \\
63 & $297$ & $\scriptstyle A_{2} + 2D_{8}$ & 0 & 512 & 196.54 \\
64 & $217$ & $\scriptstyle A_{1} + D_{8} + D_{9}$ & 256 & 416 & 148.43 \\
65 & $218$ & $\scriptstyle D_{8} + D_{10}$ & 512 & 448 & 104.94 \\
66 & $241(1)$ & $\scriptstyle A_{1} + A_{11} + E_{6}+A_1^{(\rm R)}$ & 412 & 0 & 82.53 \\
67 & $242$ & $\scriptstyle A_{12} + E_{6}+A_1^{(\rm R)}$ & 456 & 0 & 52.15 \\
68 & $255$ & $\scriptstyle D_{4} + D_{8} + E_{6}$ & 768 & 256 & 61.05 \\
69 & $255$ & $\scriptstyle A_{1} + A_{5} + 2D_{6}$ & 384 & 248 & 98.28 \\
70 & $259$ & $\scriptstyle 2A_{1} + 2A_{3} + A_{4} + D_{6}$ & 128 & 160 & 120.95 \\
71 & $290$ & $\scriptstyle A_{1} + D_{10} + E_{7}+2A_1^{(\rm R)}$ & 1232 & 0 & $-$87.21 \\
\end{longtable}
\endgroup

Finally, we may use these results to  check the refined de Sitter swampland conjecture. 
The refined de Sitter conjecture \cite{Garg:2018reu,Ooguri:2018wrx} states that, in a region with $V>0$,
\bea
\label{eq:swampland}
    \frac{|\nabla V|}{V}\geq c \qquad \text{or} \qquad  \frac{{\min}\!\left(g^{ac}\nabla_c\nabla_b V\right)}{V}  \leq -c' ~,
\eea
where $c,c'>0$ are order-one constants, $g^{ab}$ is the inverse metric on moduli space, and ${\min}$ denotes the smallest eigenvalue of the covariant Hessian with one index raised. Equivalently, it is the smallest eigenvalue of the Hessian evaluated in an orthonormal frame. 

Note that in general, the one-loop vacuum energy is independent of the dilaton in the $D$-dimensional string frame, but after transforming to the Einstein frame it takes the form
\bea
V(\phi_D,\mathcal{M})=e^{{2D \over D-2}\phi_D}\Lambda(\mathcal{M})~,
\eea
where $\phi_D$ is the $D$-dimensional dilaton and $\mathcal{M}$ denotes the Narain moduli. It follows that
\bea
\p_{\phi_D}V={2D\over D-2}V~, 
\eea
so the potential has a nonzero, order-one dilaton slope wherever $V\neq0$. Thus, even when $\Lambda$ is stationary with respect to all Narain moduli, the corresponding point is not a critical point of the full scalar potential. In this case, the first inequality of (\ref{eq:swampland}) is already satisfied by the dilaton direction, and the refined de Sitter conjecture imposes no additional requirement on the Hessian in the Narain directions. 
To formulate a test of the second inequality of the conjecture, we thus assume that some additional dynamics stabilizes the $D$-dimensional dilaton at some fixed value. More precisely, we assume that the stabilizing sector cancels the dilaton slope and gives $\phi_D$ a positive mass, while having negligible dependence on the Narain moduli and negligible vacuum energy at the stabilization point. The effective potential for the remaining light fields is then obtained by fixing $\phi_D$. At the maximally enhanced points considered below, all derivatives with respect to the Narain moduli vanish, so the first inequality of (\ref{eq:swampland}) is violated and it becomes meaningful to test whether the smallest eigenvalue of the fixed-$\phi_D$ Hessian satisfies the second inequality. This is what we check for $D=8$ below.

We list in Table \ref{tab:intro8d} the dimensionless ratios ${\min}\!\left(g^{ac}\nabla_c\nabla_b \Lambda^{(8)}\right)/\Lambda^{(8)}$ for the 25 points of maximal gauge enhancement in the eight-dimensional compactification that are tachyon-free and scalar-free at tree level; some details regarding the computation are given in Appendix \ref{app:modulispace}. From these values, we obtain an effective bound $c'=0.67$. For comparison, the corresponding value at the maximal-enhancement points in the nine-dimensional compactification is $c'=0.64$ \cite{Fraiman:2023cpa}. 

Thus far, the de Sitter swampland conjecture has its strongest general support in parametrically controlled asymptotic regions of moduli space \cite{Ooguri_2019}, near infinite-distance boundaries where towers of states become light. Away from such limits, the proposed bounds have been tested
in various explicit string compactifications, see e.g. \cite{Blanco-Pillado:2018xyn, Gonzalo:2018guu, Fraiman:2023cpa}, although the available evidence is model-dependent and no general derivation is known. Our calculation should therefore be regarded as an explicit string-theoretic test of the refined de Sitter conjecture at finite distance.

\begingroup
\footnotesize
\setlength{\tabcolsep}{2pt}
\renewcommand{\arraystretch}{1.0}

\setlength\LTleft{\fill}
\setlength\LTright{\fill}

\begin{longtable}{|c!{\vrule width 1.2pt}p{0.94\textwidth}|}
\caption{Eigenvalues of the physical Hessian for the tachyon- and scalar-free eight-dimensional entries.}
\label{tab:hessian8d}
\\
\hline
$\#$ & Eigenvalues of $g^{ac}\nabla_c \nabla_b \Lambda^{(8)}\times 10^6$ \\
\hline\hline
\endfirsthead

\multicolumn{2}{c}{\tablename\ \thetable\ continued from the previous page} \\
\hline
$\#$ & Eigenvalues of $g^{ac}\nabla_c \nabla_b \Lambda^{(8)}\times 10^6$ \\
\hline\hline
\endhead

\hline
\endfoot

\hline
\endlastfoot

2 & 
$
-71.226^{36} \allowbreak\;
$ \\
\hline
4 & 
$
-315.01^{4},\allowbreak\; 172.56^{4},\allowbreak\; 
-71.226^{28}
$ \\
\hline
5 & 
$
-403.51^{2},\allowbreak\;
192.22^{3},\allowbreak\;
-118.40^{2},\allowbreak\;
-105.65^{13},\allowbreak\;
-75.844^{13},\allowbreak\;
-33.292^{3}
$ \\
\hline
11 & 
$
-228.16^{10},\allowbreak\; 142.74^{8},\allowbreak\; 
-116.92^{10},\allowbreak\; 
-82.703^{8}
$ \\
\hline
12 & 
$
296.92^{1},\allowbreak\; -101.55^{17},\allowbreak\; 
-98.775^{17},\allowbreak\; 
-41.852^{1}
$ \\
\hline
13 & 
$
196.59^{2},\allowbreak\; -130.99^{3},\allowbreak\; 
-107.50^{7},\allowbreak\; 
-93.402^{3},\allowbreak\; 
-90.047^{7},\allowbreak\; 
-86.691^{6},\allowbreak\; 
-84.012^{6},\allowbreak\; 
-46.606^{2}
$ \\
\hline
14 & 
$
-264.58^{2},\allowbreak\; -146.61^{7},\allowbreak\; 
145.29^{4},\allowbreak\; 
-89.798^{7},\allowbreak\; 
-76.749^{5},\allowbreak\; 
-76.216^{2},\allowbreak\; 
-55.260^{5},\allowbreak\; 
-48.100^{4}
$ \\
\hline
15 & 
$
-236.05^{10},\allowbreak\; 143.35^{8},\allowbreak\; 
-89.844^{10},\allowbreak\; 
-43.499^{8}
$ \\
\hline
18 & 
$
-315.01^{6},\allowbreak\; 172.56^{6},\allowbreak\; 
-71.226^{24}
$ \\
\hline
20 & 
$
-403.51^{5},\allowbreak\; 192.22^{6},\allowbreak\; 
-118.40^{5},\allowbreak\; 
-105.65^{7},\allowbreak\; 
-75.844^{7},\allowbreak\; 
-33.292^{6}
$ \\
\hline
24 & 
$
-315.01^{10},\allowbreak\; 172.56^{10},\allowbreak\; 
-71.226^{16}
$ \\
\hline
28 & 
$
-315.01^{18},\allowbreak\; 172.56^{18}
$ \\
\hline
30 & 
$
-542.61^{5},\allowbreak\; -412.25^{4},\allowbreak\; 
-284.52^{5},\allowbreak\; 
277.96^{9},\allowbreak\; 
169.12^{4},\allowbreak\; 
-46.689^{9}
$ \\
\hline
34 & 
$
-343.56^{2},\allowbreak\; -205.63^{2},\allowbreak\; 
-176.72^{8},\allowbreak\; 
169.84^{8},\allowbreak\; 
-101.52^{16}
$ \\
\hline
40 & 
$
-252.45^{5},\allowbreak\; 228.18^{2},\allowbreak\; 
-159.84^{1},\allowbreak\; 
157.94^{4},\allowbreak\; 
-135.41^{5},\allowbreak\; 
-125.99^{1},\allowbreak\; 
-118.90^{6},\allowbreak\; 
-113.50^{6},\allowbreak\; 
-105.41^{4},\allowbreak\; 
-73.537^{2}
$ \\
\hline
41 & 
$
-269.34^{3},\allowbreak\; 235.59^{2},\allowbreak\; 
-235.18^{1},\allowbreak\; 
168.02^{4},\allowbreak\; 
-155.63^{3},\allowbreak\; 
-153.19^{2},\allowbreak\; 
-126.54^{8},\allowbreak\; 
-108.72^{8},\allowbreak\; 
-102.11^{1},\allowbreak\; 
-91.478^{4}
$ \\
\hline
42 & 
$
242.79^{4},\allowbreak\; -150.95^{4},\allowbreak\; 
-135.02^{24},\allowbreak\; 
-92.991^{4}
$ \\
\hline
43 & 
$
242.79^{4},\allowbreak\; -150.95^{8},\allowbreak\; 
-135.02^{16},\allowbreak\; 
-119.10^{4},\allowbreak\; 
-92.991^{4}
$ \\
\hline
52 & 
$
341.76^{1},\allowbreak\; -284.06^{5},\allowbreak\; 
177.55^{4},\allowbreak\; 
-135.55^{5},\allowbreak\; 
-129.44^{8},\allowbreak\; 
-126.99^{8},\allowbreak\; 
-101.79^{4},\allowbreak\; 
-63.523^{1}
$ \\
\hline
53 & 
$
-254.18^{5},\allowbreak\; -237.48^{5},\allowbreak\; 
152.66^{5},\allowbreak\; 
-132.54^{8},\allowbreak\; 
-126.44^{8},\allowbreak\; 
100.73^{5}
$ \\
\hline
56 & 
$
242.79^{4},\allowbreak\; -150.95^{16},\allowbreak\; 
-119.10^{12},\allowbreak\; 
-92.991^{4}
$ \\
\hline
57 & 
$
362.52^{1},\allowbreak\; 268.98^{2},\allowbreak\; 
-154.63^{15},\allowbreak\; 
-137.74^{15},\allowbreak\; 
-95.225^{2},\allowbreak\; 
-74.068^{1}
$ \\
\hline
58 & 
$
187.99^{3},\allowbreak\; -152.42^{15},\allowbreak\; 
-147.16^{15},\allowbreak\; 
141.96^{3}
$ \\
\hline
62 & 
$
403.69^{2},\allowbreak\; -155.02^{32},\allowbreak\; 
-83.915^{2}
$ \\
\hline
63 & 
$
199.84^{4},\allowbreak\; -155.41^{32}
$ \\
\hline
\end{longtable}
\endgroup

\section*{Acknowledgments}
Numerical computations in this work were carried out on the Genkai cluster at Kyushu University. JK thanks Savdeep Sethi for useful correspondences.  YH is supported by Grant-in-Aid for JSPS Fellows No. 25KJ1925. JK is supported by JSPS KAKENHI Grant No. 26K17152 and through the Inamori Frontier Program at Kyushu University.

\appendix

\section{Ten-dimensional orbifolds}
\label{app:10d}

In this appendix, we review the well known results of \cite{Dixon:1986iz}, which used the orbifold construction to identify ten-dimensional non-supersymmetric string theories. In the process, we also set our conventions for the various ingredients appearing in the Kac construction.

 Note that with no geometric torus, the Narain vector specifying the orbifold reduces to a gauge-lattice shift $\delta_{16}$.  The consistency conditions (\ref{eq:deltaconstraints}) then become
\bea
\label{eq:10dconsistency}
   2\delta_{16}\in\Gamma_{16}~, \hspace{0.4 in}  \delta_{16}^{2}\in\mathbb Z~,
\eea
and the invariant roots are those with integral inner product with $\delta_{16}$. 

\subsection{$E_8\times E_8$ string}
    \begin{figure}[ht]
    \centering
    \begin{tikzpicture}[scale=0.7]
    \pgfmathsetmacro{\myxlow}{-1}
    \pgfmathsetmacro{\myxhigh}{1}

    \node[thick, circle, draw, fill=white] (1) at (0,0) {};
    \node[thick, circle, draw, fill=white] (2) at (1,0) {};
    \node[thick, circle, draw, fill=white] (3) at (2,0) {};
    \node[thick, circle, draw, fill=white] (4) at (3,0) {};
    \node[thick, circle, draw, fill=white] (5) at (4,0) {};
    \node[thick, circle, draw, fill=white] (6) at (5,0) {};
    \node[thick, circle, draw, fill=white] (7) at (1,1) {};
    \node[thick, circle, draw, fill=white] (8) at (1,2) {};
    \node[thick, circle, draw, fill=yellow] (0) at (6,0) {};

    \draw[thick] (1) -- (2);
    \draw[thick] (2) -- (3);
    \draw[thick] (3) -- (4);
    \draw[thick] (4) -- (5);
    \draw[thick] (5) -- (6);
    \draw[thick] (2) -- (7);
    \draw[thick] (7) -- (8);
    \draw[thick] (6) -- (0);
    \node[below=0.3cm] at (1) {1};
    \node[below=0.3cm] at (2) {2};
    \node[below=0.3cm] at (3) {3};
    \node[below=0.3cm] at (4) {4};
    \node[below=0.3cm] at (5) {5};
    \node[below=0.3cm] at (6) {6};
    \node[left=0.3cm]  at (7) {7};
    \node[left=0.3cm]  at (8) {8};
    \node[below=0.3cm] at (0) {0};
    \node[below=0.3cm, yshift=-12pt, text=red] at (1) {3};
    \node[below=0.3cm, yshift=-12pt, text=red] at (2) {6};
    \node[below=0.3cm, yshift=-12pt, text=red] at (3) {5};
    \node[below=0.3cm, yshift=-12pt, text=red] at (4) {4};
    \node[below=0.3cm, yshift=-12pt, text=red] at (5) {3};
    \node[below=0.3cm, yshift=-12pt, text=red] at (6) {2};
    \node[below=0.3cm, yshift=-12pt, text=red] at (0) {1};
    \node[left=0.3cm, xshift=-10pt, text=red] at (7) {4};
    \node[left=0.3cm, xshift=-10pt, text=red] at (8) {2};
    \end{tikzpicture}
    \caption{Affine Dynkin diagram of ${E}_8$. The Kac marks are shown in red.}
    \label{Dynkin_affineE8}
    \end{figure}
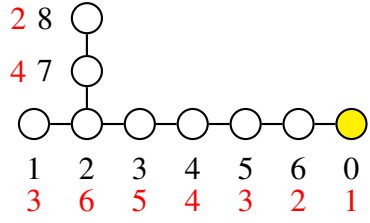

We begin with the supersymmetric $(E_8 \times E_8)\rtimes \ZZ_2$ heterotic string. For each $E_8$ factor, order-two inner automorphisms can be read from the affine diagram in Figure \ref{Dynkin_affineE8}.  The simple roots, fundamental weights, and Kac marks in our basis are collected in Table \ref{tab:E8_simple_roots}.  Because the $E_8$ weight and root lattices coincide, the Kac weight already lies in a particularly simple ambient lattice; the nontrivial restriction comes from imposing (\ref{eq:10dconsistency}) on the pair of factors.

\begin{table}[h]
    \centering
    \begin{tabular}{|c|c||c|c|} \hline
        $~i~$ & $\kappa_i$ & $\alpha_i$ & $\omega_i$ \\ \hline
        1 & $3$ & $(1,\minus1,0,0,0,0,0,0)$ & $\minus(\minus\frac{1}{2},\frac{1}{2},\frac{1}{2},\frac{1}{2},\frac{1}{2},\frac{1}{2},\frac{1}{2},\minus\frac{5}{2})$ \\
        2 & $6$ & $(0,1,\minus1,0,0,0,0,0)$ & $\minus(0,0,1,1,1,1,1,\minus5)$ \\
        3 & $5$ & $(0,0,1,\minus1,0,0,0,0)$ & $\minus(0,0,0,1,1,1,1,\minus4)$ \\
        4 & $4$ & $(0,0,0,1,\minus1,0,0,0)$ & $\minus(0,0,0,0,1,1,1,\minus3)$ \\
        5 & $3$ & $(0,0,0,0,1,\minus1,0,0)$ & $\minus(0,0,0,0,0,1,1,\minus2)$ \\
        6 & $2$ & $(0,0,0,0,0,1,\minus1,0)$ & $(0,0,0,0,0,0,\minus1,1)$ \\
        7 & $4$ & $\minus(1,1,0,0,0,0,0,0)$ & $\minus(\frac{1}{2},\frac{1}{2},\frac{1}{2},\frac{1}{2},\frac{1}{2},\frac{1}{2},\frac{1}{2},\minus\frac{7}{2})$ \\
        8 & $2$ & $(\frac{1}{2},\frac{1}{2},\frac{1}{2},\frac{1}{2},\frac{1}{2},\frac{1}{2},\frac{1}{2},\frac{1}{2})$ & $(0,0,0,0,0,0,0,2)$ \\
        0 & $1$ & $(0,0,0,0,0,0,1,\minus1)$ & $(0,0,0,0,0,0,0,0)$ \\ \hline
    \end{tabular}
    \caption{Kac marks, simple roots, and fundamental weights of $E_8$.}
    \label{tab:E8_simple_roots}
\end{table}

For the ten-dimensional $E_8\times E_8$ string, the Kac data may be chosen independently in the two factors, subject to the lattice norm condition.  We denote the two sets of affine labels by $s_i$ and $s'_i$,
\bea
    \mathbf{s}=(s_1,s_2,s_3,s_4,s_5,s_6,s_7,s_8,s_0;s_1',s_2',s_3',s_4',s_5',s_6',s_7',s_8',s_0')~,
\eea
which determine the sixteen-dimensional shift $\delta=\delta_8\oplus\delta'_8$.  Applying (\ref{eq:10dconsistency}) leaves the following four representatives,
\bea
    \mathbf{s}=(0^7,1,0;0^7,1,0),~(0^5,1,0^3;0^5,1,0^3),~(0^8,2;0^7,1,0),~(0^7,1,0;0^8,2)~,
\eea
which give
\bea
    \delta=(0^7,1;0^7,1),~\left(0^6,-\frac{1}{2},\frac{1}{2};0^6,-\frac{1}{2},\frac{1}{2}\right),~(0^8;0^7,1),~(0^7,1;0^8)~.
\eea
The corresponding invariant algebras are $2D_8$, $2E_7+2A_1$, and $D_8+E_8$.  The last two vectors are exchanged by interchanging the two $E_8$ factors and hence yield the same abstract ten-dimensional theory.  

\subsection{$\mathrm{Spin}(32)/\mathbb{Z}_2$ parent}

\begin{figure}[h]
    \centering
    \begin{tikzpicture}[scale=0.7]
    \pgfmathsetmacro{\myxlow}{-1}
    \pgfmathsetmacro{\myxhigh}{1}
    \node[thick, circle, draw, fill=white] (1) at (0,0) {};
    \node[thick, circle, draw, fill=white] (2) at (1,0) {};
    \node[thick, circle, draw, fill=white] (3) at (2,0) {};
    \node[thick, circle, draw, fill=white] (4) at (3,0) {};
    \node[thick, circle, draw, fill=white] (5) at (4,0) {};
    \node[thick, circle, draw, fill=white] (6) at (5,0) {};
    \node[thick, circle, draw, fill=white] (7) at (6,0) {};
    \node[thick, circle, draw, fill=white] (8) at (7,0) {};
    \node[thick, circle, draw, fill=white] (9) at (8,0) {};
    \node[thick, circle, draw, fill=white] (10) at (9,0) {};
    \node[thick, circle, draw, fill=white] (11) at (10,0) {};
    \node[thick, circle, draw, fill=white] (12) at (11,0) {};
    \node[thick, circle, draw, fill=white] (13) at (12,0) {};
    \node[thick, circle, draw, fill=white] (14) at (13,0) {};
    \node[thick, circle, draw, fill=white] (15) at (14,0) {};
    \node[thick, circle, draw, fill=white] (16) at (13,1) {};
    \node[thick, circle, draw, fill=yellow] (17) at (1,1) {};
    
    \draw[thick] (1) -- (2);
    \draw[thick] (2) -- (3);
    \draw[thick] (3) -- (4);
    \draw[thick] (4) -- (5);
    \draw[thick] (5) -- (6);
    \draw[thick] (6) -- (7);
    \draw[thick] (7) -- (8);
    \draw[thick] (8) -- (9);
    \draw[thick] (9) -- (10);
    \draw[thick] (10) -- (11);
    \draw[thick] (11) -- (12);
    \draw[thick] (12) -- (13);
    \draw[thick] (13) -- (14);
    \draw[thick] (14) -- (15);
    \draw[thick] (14) -- (16);
    \draw[thick] (2) -- (17);

    \node[below=0.3cm] at (1) {1};
    \node[below=0.3cm] at (2) {2};
    \node[below=0.3cm] at (3) {3};
    \node[below=0.3cm] at (4) {4};
    \node[below=0.3cm] at (5) {5};
    \node[below=0.3cm] at (6) {6};
    \node[below=0.3cm] at (7) {7};
    \node[below=0.3cm] at (8) {8};
    \node[below=0.3cm] at (9) {9};
    \node[below=0.3cm] at (10) {10};
    \node[below=0.3cm] at (11) {11};
    \node[below=0.3cm] at (12) {12};
    \node[below=0.3cm] at (13) {13};
    \node[below=0.3cm] at (14) {14};
    \node[below=0.3cm] at (15) {15};
    \node[right=0.3cm] at (16) {16};
    \node[left=0.3cm] at (17) {17};
    \node[below=0.3cm, yshift=-15pt, text=red] at (1) {1};
    \node[below=0.3cm, yshift=-15pt, text=red] at (2) {2};
    \node[below=0.3cm, yshift=-15pt, text=red] at (3) {2};
    \node[below=0.3cm, yshift=-15pt, text=red] at (4) {2};
    \node[below=0.3cm, yshift=-15pt, text=red] at (5) {2};
    \node[below=0.3cm, yshift=-15pt, text=red] at (6) {2};
    \node[below=0.3cm, yshift=-15pt, text=red] at (7) {2};
    \node[below=0.3cm, yshift=-15pt, text=red] at (8) {2};
    \node[below=0.3cm, yshift=-15pt, text=red] at (9) {2};
    \node[below=0.3cm, yshift=-15pt, text=red] at (10) {2};
    \node[below=0.3cm, yshift=-15pt, text=red] at (11) {2};
    \node[below=0.3cm, yshift=-15pt, text=red] at (12) {2};
    \node[below=0.3cm, yshift=-15pt, text=red] at (13) {2};
    \node[below=0.3cm, yshift=-15pt, text=red] at (14) {2};
    \node[below=0.3cm, yshift=-15pt, text=red] at (15) {1};
    \node[right=0.3cm, xshift=15pt, text=red] at (16) {1};
    \node[left=0.3cm, xshift=-15pt, text=red] at (17) {1};
    \end{tikzpicture}
    \caption{Affine Dynkin diagram of ${D}_{16}$. The Kac marks are shown in red.}
    \label{Dynkin_affineD16}
    \end{figure}
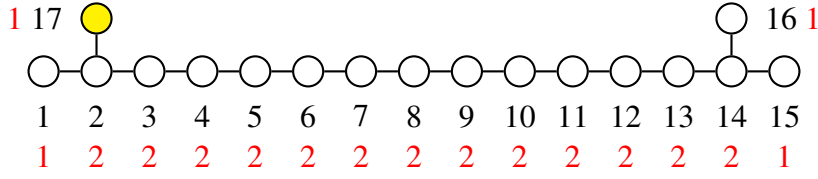

For the $\mathrm{Spin}(32)/\mathbb Z_2$ parent, the gauge lattice is the even self-dual extension $D_{16}^{+}$ rather than the $D_{16}$ root lattice alone.  
We use the following simple roots and fundamental weights,
\bea
    &\beta_{k}=(0^{k-1},1,-1,0^{15-k})~,\qquad \mathrm{w}_{k}=(1^{k},0^{16-k})~,\qquad k=1,\ldots,14\\
    &\beta_{15}=(0^{14},1,-1)~,\qquad \mathrm{w}_{15}=\left(\left(\frac{1}{2}\right)^{15},-\frac{1}{2}\right)~,\\
    &\beta_{16}=(0^{14},1,1)~,\qquad \mathrm{w}_{16}=\left(\left(\frac{1}{2}\right)^{16}\right)~.
\eea
The lowest root is $\beta_{17}=(-1,-1,0^{14})$, and the Kac marks are shown in Figure~\ref{Dynkin_affineD16}.

For the ten-dimensional $\mathrm{Spin}(32)/\mathbb Z_2$ string, introduce non-negative integers $\widetilde s_i$ for $i=1,\ldots,17$, and define the order and shift by
\bea
   \tilde{N}:=\sum_{i=1}^{17}\tilde{s}_{i}\kappa_{i}~, \hspace{0.4 in}\tilde{\delta}:=\frac{1}{\tilde{N}}\sum_{i=1}^{16}\tilde{s}_{i}\mathrm{w}_{i}~.
\eea
Primitive labels describe an adjoint automorphism of exact order $\widetilde N$; as described in Section \ref{sec:Kactheorem}, non-primitive labels are retained when the full lattice character is still nontrivial.  Membership of $\widetilde N\widetilde\delta$ in $D_{16}^{+}$ requires the additional parity condition
\bea
    \sum_{\substack{1\leq i\leq15\\ i\ {\rm odd}}}\tilde{s}_{i}\in2\mathbb{Z}~.
\eea
The maximally semisimple order-two solutions of (\ref{eq:10dconsistency}) are represented by
\bea
    (\tilde{s}_1,\ldots,\tilde{s}_{17})=(0^7,1,0^9),~(0^3,1,0^{13}),~(0^{11},1,0^5),~(2,0^{16}),~(0^{14},2,0^2)~,
\eea
which give
\bea
    \tilde{\delta}=\left(\left(\frac{1}{2}\right)^8,0^8\right),~\left(\left(\frac{1}{2}\right)^4,0^{12}\right),~\left(\left(\frac{1}{2}\right)^{12},0^{4}\right),~(1,0^{15}),~\left(\left(\frac{1}{2}\right)^{15},-\frac{1}{2}\right)~.
\eea
They give the invariant algebras $2D_8$, $D_4+D_{12}$, and $D_{16}$.  The second and third shifts are related by exchanging the complementary blocks of four and twelve directions, while the last two act trivially on the $D_{16}$ roots but nontrivially on the glue vectors of $D_{16}^{+}$.

For comparison, a representative with two nonzero mark-one labels is
\bea
(\tilde s_1,\ldots,\tilde s_{17})=(0^{15},1,1)~,  \hspace{0.4 in}\tilde\delta=\left(\left(\frac14\right)^{16}\right)~.
\eea
It satisfies $2\tilde\delta\in D_{16}^{+}$ and $\tilde\delta^2=1$.  The roots $e_i-e_j$ survive, whereas $e_i+e_j$ are projected out, so the invariant algebra is $A_{15}+\mathfrak u(1)$.  This is the ten-dimensional $U(16)$ branch.  We display it only to emphasize the scope of our classification: deleting two affine nodes lowers the semisimple rank by one and leaves a Cartan current as an explicit left-moving Abelian factor.  Such shifts are excluded from the maximally semisimple eight-dimensional tables.

 \section{Details on the Hessian computation}
 \label{app:modulispace}
In order to properly compute the Hessian in (\ref{eq:hessian}), it is necessary for us to use the metric on moduli space $\Gamma_{d,16+d}$, which can be found in \cite{Maharana:1992my}. Below we review these results and write the explicit form of the metric for the case of $d=2$. 

 \subsection{$T^d$ with trivial Wilson lines}

Consider the compactification of the ten-dimensional heterotic string on a torus $T^d$, with  $D = 10-d$ non-compact spacetime dimensions. The ten-dimensional string-frame action is given by
\bea
\label{eq:originalheteroticaction}
  S_{10}  = {1 \over 2\kappa_{10}^2} \int d^{10}x\,\sqrt{-G^{(10)}_{s}}\,e^{-2\Phi}  \left[ R[G^{(10)}_s] +4(\p\Phi)^2 -{1\over 12}H_{MNP}H^{MNP} -{1 \over 4}F^I_{MN}F^{I,MN}+\cdots \right] ~
  \no\\
\eea
for $\alpha'=1$; the precise overall normalization will not play any role.  The internal metric is $G_{ij}(x)$, where $i,j=1,\ldots,n$.  We ignore graviphotons in the expressions below, since they do not affect the scalar sigma-model metric at the order studied here. We take as our metric Ansatz the following,
\bea
\label{eq:metricAnsatz}
d s^2_{10,s} = g^{(D),s}_{\mu\nu}(x)\,d x^\mu d  x^\nu +G_{ij}(x)\,d y^i d  y^j ~,
\eea
with $x^\mu$ the $D$-dimensional coordinates and $y^i$ the coordinates of $T^d$.  Define the internal volume and its logarithm by
\bea
  \mathrm{vol}(T^n)=\sqrt{\det G_{ij}}~, \hspace{0.4 in} v := \mathrm{log}\, \mathrm{vol}(T^n)~. 
\eea
The $D$-dimensional string-frame dilaton is
\bea
  \phi_D = \Phi-{1\over 2}v  ~,
\eea
as follows from
\bea
  \sqrt{-G^{(10)}_{s}}\,e^{-2\Phi}  =\sqrt{- g^{(D),s}}\,\mathrm{vol}(T^d)\,e^{-2\Phi} =\sqrt{- g^{(D),s}}\,e^{-2\phi_D} ~. 
\eea
The $D$-dimensional string- and Einstein-frame metrics are related by
\bea
  g^{(D),s}_{\mu\nu} =e^{{4 \over D-2}\phi_D}\,g^{(D),E}_{\mu\nu} ~,
\eea
or equivalently, in terms of the ten-dimensional dilaton and volume,
\bea
  g^{(D),s}_{\mu\nu}  =e^{{4\over D-2}\Phi} e^{-{2\over D-2}v} g^{(D),E}_{\mu\nu} ~.
\eea

Now for the metric Ansatz in (\ref{eq:metricAnsatz}), the ten-dimensional Ricci scalar may be expanded as, 
\bea
R[G^{(10)}_{s}] = R[g^{(D)}_s] - 2 \nabla^2 v - (\p v)^2 + {1\over 4} \mathrm{Tr}(\p G^{-1} \p G)~, 
\eea
where contractions are performed using the $D$-dimensional string-frame metric. 
Performing an integration by parts on the second term, 
\bea
\int \sqrt{- g^{(D)}_s}\, e^{- 2 \phi_D} (-2 \nabla^2 v) = - 4 \int \sqrt{- g^{(D)}_s}\, e^{- 2 \phi_D}\, \p \phi_D \p v~,
\eea
and adding in the dilaton kinetic term from (\ref{eq:originalheteroticaction}), we find 
\bea
\sqrt{-G^{(10)}_{s}}\,e^{-2\Phi}  \left[ R[G^{(10)}_s] +4(\p\Phi)^2 \right]  &=& \sqrt{- g^{(D)}_s}\, e^{- 2 \phi_d}\left[R[g^{(D)}_s]  + 4 (\p\phi_D)^2 + {1\over 4}\mathrm{Tr}(\p G^{-1} \p G) \right] 
\no\\
&=&  \sqrt{- g^{(D)}_E}\, \left[R[g^{(D)}_E]  - {4 \over D - 2} (\p\phi_D)^2 + {1\over 4}\mathrm{Tr}(\p G^{-1} \p G) \right] ~.
\no\\
\eea
On the other hand, the internal components of the two-form and gauge fields give scalar fields $B_{ij}(x)$ and $A_i^I(x)$ with action 
\bea
- {1\over 4}G^{ik} G^{j\ell} \p B_{ij} \p B_{k \ell} - {1 \over 2} G^{ij} \p A^I_i \p A^I_j~
\eea
where for simplicity we begin with the case of zero Wilson lines. 
In total, the scalar kinetic terms are as follows,
\bea
- {4 \over D-2} \left( \p \Phi - \half \p v\right)^2 + {1\over 4}\mathrm{Tr}(\p G^{-1} \p G) - {1\over 4}G^{ik} G^{j\ell} \p B_{ij} \p B_{k \ell} - {1 \over 2} G^{ij} \p A^I_i \p A^I_j~.
\eea
The metric on moduli space can then be written as follows,
\bea
\label{eq:metriconmodulispace}
ds^2_\mathrm{moduli} = {8 \over D - 2}\left(d \Phi- \half dv \right)^2 - \half \mathrm{Tr} (d G^{-1} d G) + \half G^{ik} G^{j \ell} d B_{ij} d B_{k \ell} +  G^{ij} d A^I_i d A_j^I~. 
\no\\
\eea

\subsection{$T^2$ with trivial Wilson lines}

In the case of $T^2$ we have complex structure and K{\"a}hler moduli $\tau$ and $\rho$, related to the metric and $B$-field via \cite{Giveon:1994fu}
\bea
G_{ij} = {\rho_2 \over \tau_2} \left( \begin{matrix} |\tau|^2 & \tau_1 \\ \tau_1 & 1 \end{matrix} \right) ~, 
\hspace{0.4 in}
B_{ij} = \left( \begin{matrix} 0 & \rho_1 \\ -\rho_1 & 0 \end{matrix} \right) ~. 
\eea
Note that 
\bea
\mathrm{vol}(T^n) = \sqrt{\mathrm{det} G} = \rho_2~.
\eea
We may now compute the terms in  (\ref{eq:metriconmodulispace}) explicitly. We take $\phi_8$ to be held fixed, in which case the first term vanishes. 
On the other hand, the second and third terms are
\bea
 - \half \mathrm{Tr} (d G^{-1} d G)  = {d \rho_2^2 \over \rho_2^2} + {d \tau_1^2 + d \tau_2^2 \over \tau_2^2}~, \hspace{0.4 in}\half G^{ik} G^{j \ell} d B_{ij} d B_{k \ell} = {d \rho_1^2 \over \rho_2^2} ~. 
\eea
Putting these together, the final metric on moduli space for fixed $\Phi$ is
\bea
ds^2_\mathrm{moduli} ={d \rho_1^2 \over \rho_2^2}+{d \rho_2^2 \over \rho_2^2} + {d \tau_1^2 + d \tau_2^2 \over \tau_2^2}  + G^{ij} d A^I_i d A_j^I~. 
\eea
So the Hessian should be scaled as follows for the metric moduli
\bea
\label{eq:Hessianattempt1}
H_{\rho_1 \rho_1} = \rho_2^2\, \p_{\rho_1}^2 \Lambda~, \hspace{0.4 in} H_{\rho_2 \rho_2} =\rho_2^2\, \p_{\rho_2}^2 \Lambda~, \hspace{0.4 in} H_{\tau_i \tau_i} = \tau_2^2\, \p_{\tau_i} \p_{\tau_i} \Lambda~.
\eea
For the remaining moduli things are more complicated, since we have non-trivial mixing. We may simplify things by switching to the following variables, 
\bea
W^I := A_1^I - \tau A_2^I~, 
\eea
for which we can rewrite 
\bea
G^{ij} d A^I_i d A_j^I = {| d W^I|^2 \over \rho_2 \tau_2}~
\eea
when evaluated around a background with zero Wilson line $A^I=0$. 
Splitting $W^I$ into real and imaginary parts $W^I = W^I_1 + i W^I_2$, the variables $W^I_{1,2}$ give the natural coordinates on moduli space, and the Hessian is correspondingly given by
\bea
H_{W^I_i, W^J_j} = {\rho_2 \tau_2 }\, \p_{W^I_i} \p_{W^J_j} \Lambda~.
\eea
The normalizations of mixed entries follow similarly. 

 \subsection{Including non-trivial Wilson lines}
 
 To incorporate non-trivial Wilson lines, we should use the full metric on $O(d,d+16)$ moduli space.  This was derived in \cite[Section 4]{Maharana:1992my}.  
 We may summarize the discussion as follows. Let $A$ be the $16 \times d$ matrix with entries $A_i^I$ and define 
 \bea
 C : = B + {1 \over 2} A^T A~. 
 \eea
 Then we define the following $O(d,d+16)$ matrix 
 \bea
 M = \left(\begin{matrix} 
 G^{-1} & - G^{-1} C & - G^{-1} A^T \\
 -C^T G^{-1} & G + C^T G^{-1} C + A^T A & C^T G^{-1} A^T + A^T \\
 - A G^{-1} & A G^{-1} C + A & \mathds{1}_{16} + A G^{-1} A^T  \end{matrix}  \right) ~,
 \eea
 in terms of which the kinetic terms for the scalars can be written as\footnote{Note that a simple way to compute $M^{-1}$ is to define the matrix 
 \bea
 \eta = \left( \begin{matrix} 0 & \mathds{1}_d & 0 \\ \mathds{1}_d & 0 & 0 \\ 0 & 0 & \mathds{1}_{16}\end{matrix}  \right) ~,
 \eea
 in terms of which we may write $M^{-1} = \eta M \eta$.
 }  
 \bea
 \cL_\mathrm{kin} = {1\over 8} \mathrm{Tr} \left(\p_\m M^{-1} \p^\mu M \right)~.
 \eea
 Adding in the volume term identified previously, we thus have 
 \bea
 ds_\mathrm{moduli}^2 ={8 \over D - 2}\left(d \Phi- \half dv \right)^2  - {1\over 4}\mathrm{Tr} \left(d M^{-1} d M \right)~.
 \eea
 It is easy to check that for $A=0$, this reduces to the expression in (\ref{eq:metriconmodulispace}). 
 
 Now in terms of this metric, the physical Hessian is  given by 
 \bea
 {H^a}_b &=& g^{ac} H_{cb} = g^{ac} \nabla_c \nabla_b \Lambda
 \\\no
 &=&g^{ac} \left (\p_c \p_b \Lambda - \Gamma^d_{cb} \p_d \Lambda\right) ~,
 \eea
 where $ g^{ac}$ is the inverse metric on moduli space and $\Gamma^a_{bc}$ are the Christoffel connections constructed from it. This is the matrix whose eigenvalues we are interested in.  
 If we are focused on extrema, then we can drop the latter piece, and we have 
  \bea
  \label{eq:hessianfinal}
 {H^a}_b &=&  g^{ac} \p_c \p_b \Lambda~. 
 \eea
 
In fact, the maximal enhancement points we are studying are expected to be extrema \cite{Ginsparg:1986wr}. Indeed, at an enhancement point, the invariant roots $\alpha$ of $\Ln$ satisfy $\alpha \cdot \delta \in \ZZ$, which means that the shift character is invariant under a Weyl reflection 
 \bea
 s_\alpha(\delta) := \delta - (\alpha \cdot \delta)\alpha~, 
 \eea
 since $ s_\alpha(\delta)$ and $\delta$ differ by a Narain lattice vector. This means that the orbifold partition function, and hence the one-loop potential, is invariant under the Weyl group of $\Ln$. 
 Now the tangent space to the Narain moduli is generically of the form
 \bea
 T\cM_\mathrm{Narain} \cong \Pi_L \otimes \Pi_R^*~, 
 \eea
 and at an enhancement point the left-moving plane is identified with the Cartan subalgebra $\mathfrak{h}_L$ of the enhanced left-moving algebra. Thus the cotangent space may be written as
  \bea
 T^*\cM_\mathrm{Narain} \cong \mathfrak{h}_L^* \otimes \Pi_R~.
 \eea
 The Weyl group acts on $ \mathfrak{h}_L^*$ and leaves $\Pi_R$ invariant. If $\Ln$ is a maximal semi-simple enhancement point, its roots span $ \mathfrak{h}_L^*$, so the only covector fixed by all Weyl reflections is zero, and hence there is no Weyl-invariant one-form on the Narain moduli space. Since $d\Lambda$ is Weyl-invariant, we conclude that $d\Lambda = 0$ along Narain moduli space directions. Thus maximally semisimple enhancement points should be stationary points, and we may use the simplified expression for the Hessian (\ref{eq:hessianfinal}).

To conclude, the physical Hessian should include a multiplication by the inverse metric, which for $d=2$ and fixed $\phi_8$ takes the form 
 \bea
g^{ab} = \left(\begin{matrix}
\gamma & 0 & 0 & 0 & \alpha_J & \beta_J 
\\
0 & \rho_2^2 & 0 & 0 & 0 & 0 
\\
0 & 0 & \tau_2^2 & 0 & 0 & 0 
\\
0 & 0 & 0 & \tau_2^2 & 0 & 0 
\\ 
\alpha_I & 0 & 0 & 0& {\rho_2  |\tau|^2\over  \tau_2} \delta_{IJ} & {\rho_2 \tau_1 \over  \tau_2} \delta_{IJ}
\\
\beta_I & 0 & 0 & 0 &  {\rho_2 \tau_1 \over  \tau_2} \delta_{IJ} & {\rho_2 \over  \tau_2} \delta_{IJ}
\end{matrix} \right) ~,
\eea
 where the last two rows/columns are $16\times 16$, and we have defined 
 \bea
 \alpha_I &:=& {\rho_2 \over 2 \tau_2 }( |\tau|^2 A_2^I - \tau_1 A_1^I)~, \hspace{0.4 in} \beta_I:= {\rho_2 \over 2 \tau_2} (\tau_1 A_2^I -  A_1^I)~, 
 \no\\
 \gamma &:=& \rho_2^2 + { \rho_2 \over 4 \tau_2} (|A_1|^2 - 2 \tau_1 A_1 \cdot A_2 +|\tau|^2 |A_2|^2)~.
 \eea
Note that when $A^I = 0$ we reduce to the results in (\ref{eq:Hessianattempt1}).

\subsection{Weyl constraints on the Hessian spectrum}\label{app:Hessian spectrum}

The Weyl invariance discussed above also gives a stringent check of the numerically computed Hessian.  
Consider any of the maximally enhanced points, and decompose its semisimple left-moving algebra and Weyl group as
\bea
    \mathfrak{g} =  \bigoplus_{s=1}^{m}\mathfrak{g}_s~, \hspace{0.4 in} r_s: =\operatorname{rank}\mathfrak{g}_s~, \hspace{0.4 in}   \mathcal{W}(\mathfrak g)=\prod_{s=1}^{m}\mathcal{W}_s ~. 
\eea
Here $s$ labels each occurrence of a simple factor, so that two isomorphic
factors are counted separately.

Let $V_s$ denote the real reflection representation of $\mathcal{W}_s$, which may be identified with $\mathfrak h_s$ or $\mathfrak h_s^*$ using the invariant metric.  Since the right-moving plane $\Pi_R$ is two-dimensional and is fixed by the left-moving Weyl group, the tangent space at the maximally enhanced point decomposes as
\bea
    T\mathcal M_{\rm Narain} \cong \bigoplus_{s=1}^{m} \left(V_s\otimes\Pi_R\right)~, \hspace{0.4 in} \dim V_s=r_s~, \hspace{0.4 in}\dim\Pi_R=2~. 
\eea
The enhanced point is fixed by $\mathcal{W}(\mathfrak g)$, while both the one-loop potential and the Narain metric are Weyl invariant.  It follows that the Hessian commutes with the Weyl action,
\bea
    \rho(w)\, H\,\rho(w)^{-1} = H~, \hspace{0.4 in}w\in \mathcal{W}(\mathfrak g)~. 
\eea

Note that there can be no Hessian block mixing two distinct simple factors. Indeed, for $s\neq t$, invariance under $\mathcal{W}_s$, which acts trivially on $V_t\otimes\Pi_R$, would require the image of such a block to lie in
\bea
    V_s^{\mathcal{W}_s}\otimes\Pi_R=0~.
\eea
Moreover, the reflection representation of an irreducible Weyl group is absolutely irreducible over $\RR$.  Schur's lemma therefore gives
\bea
    \mathrm{End}_{\mathcal{W}_s}(V_s) = \RR\, \mathds{1}_{r_s}~.
\eea
Consequently, the Hessian must take the block form
\bea
    H = \bigoplus_{s=1}^{m} \left(\mathds{1}_{r_s}\otimes K_s\right)~,
\eea
where each $K_s$ is a real $2\times2$ matrix acting on $\Pi_R$.  Since $H$ is self-adjoint with respect to the moduli-space metric, the two eigenvalues of $K_s$ are real, and each occurs in the full Hessian spectrum with multiplicity $r_s$.  Equivalently, the characteristic polynomial factorizes as
\bea
    \det\!\left(\lambda\, \mathds{1}- H\right) = \prod_{s=1}^{m}  \left[\det\!\left(\lambda\,\mathds{ 1}_2-K_s\right) \right]^{r_s}~. 
\eea
As such, for each factor $\mathfrak{g}_s$, we obtain an eigenvalue of multiplicity $2 r_s$. If eigenvalues belonging to different blocks happen to coincide, these multiplicities can merge, but an elementary multiplicity cannot split. 

\begin{table}[t]
\centering
\scriptsize
\setlength{\tabcolsep}{4pt}
\renewcommand{\arraystretch}{1.08}
\begin{tabular}{c l l c}
\hline
$\#$ & $\Ln$ & multiplicities &
\\
\hline
2  & $6A_{3}$
   & $3^{ 12}$ \\
4  & $4A_{1}+2A_{7}$
   & $1^{ 8},\,7^{ 4}$\\
5  & $2A_{1}+3A_{3}+A_{7}$
   & $1^{ 4},\,3^{ 6},\,7^{ 2}$ \\
11 & $2A_{4}+2D_{5}$
   & $4^{ 4},\,5^{ 4}$\\
12 & $A_{1}+A_{7}+2D_{5}$
   & $1^{ 2},\,5^{ 4},\,7^{ 2}$ \\
13 & $A_{1}+A_{2}+A_{3}+A_{7}+D_{5}$
   & $1^{ 2},\,2^{ 2},\,3^{ 2},
      \,5^{ 2},\,7^{ 2}$ \\
14 & $2A_{1}+A_{4}+A_{7}+D_{5}$
   & $1^{ 4},\,4^{ 2},\,5^{ 2},
      \,7^{ 2}$\\
15 & $A_{8}+2D_{5}$
   & $5^{ 4},\,8^{ 2}$ \\
18 & $2A_{3}+2D_{6}$
   & $3^{ 4},\,6^{ 4}$\\
20 & $A_{7}+D_{5}+D_{6}$
   & $5^{ 2},\,6^{ 2},\,7^{ 2}$\\
24 & $2D_{5}+D_{8}$
   & $5^{ 4},\,8^{ 2}$ \\
28 & $2D_{9}$
   & $9^{ 4}$\\
30 & $A_{4}+D_{5}+D_{9}$
   & $4^{ 2},\,5^{ 2},\,9^{ 2}$\\
34 & $2A_{1}+2A_{4}+D_{8}$
   & $1^{ 4},\,4^{ 4},\,8^{ 2}$\\
40 & $A_{1}+A_{2}+A_{4}+D_{5}+D_{6}$
   & $1^{ 2},\,2^{ 2},\,4^{ 2},
      \,5^{ 2},\,6^{ 2}$ \\
41 & $A_{1}+A_{2}+A_{3}+A_{4}+D_{8}$
   & $1^{ 2},\,2^{ 2},\,3^{ 2},
      \,4^{ 2},\,8^{ 2}$ \\
42 & $2A_{1}+2A_{2}+2D_{6}$
   & $1^{ 4},\,2^{ 4},\,6^{ 4}$ \\
43 & $2A_{1}+2A_{2}+D_{4}+D_{8}$
   & $1^{ 4},\,2^{ 4},\,4^{ 2},
      \,8^{ 2}$ \\
52 & $A_{1}+A_{4}+D_{5}+D_{8}$
   & $1^{ 2},\,4^{ 2},\,5^{ 2},
      \,8^{ 2}$ \\
53 & $A_{5}+D_{5}+D_{8}$
   & $5^{ 4},\,8^{ 2}$\\
56 & $2A_{2}+D_{6}+D_{8}$
   & $2^{ 4},\,6^{ 2},\,8^{ 2}$ \\
57 & $2A_{1}+A_{2}+D_{6}+D_{8}$
   & $1^{ 4},\,2^{ 2},\,6^{ 2},
      \,8^{ 2}$ \\
58 & $A_{1}+A_{3}+D_{6}+D_{8}$
   & $1^{ 2},\,3^{ 2},\,6^{ 2},
      \,8^{ 2}$ \\
62 & $2A_{1}+2D_{8}$
   & $1^{ 4},\,8^{ 4}$ \\
63 & $A_{2}+2D_{8}$
   & $2^{ 2},\,8^{ 4}$ \\
\hline
\end{tabular}
\caption{
Elementary eigenvalue multiplicity packets implied by Weyl invariance.
}
\label{tab:weylpackets}
\end{table}

Table \ref{tab:weylpackets} lists the elementary multiplicities for all tachyon-free, scalar-free cases.  The notation $r^{ n}$ denotes $n$ elementary packets of size $r$. We used these constraints as an internal check of the numerical calculation.  Indeed, every resulting multiplicity pattern in Table \ref{tab:hessian8d} is compatible with Table \ref{tab:weylpackets}.  For example, for No. 2, Weyl invariance requires twelve packets of size three, and the numerical result shows that all twelve packets merge into a single eigenvalue of multiplicity $36$.  This additional coincidence is stronger than what follows from Weyl invariance alone.

Likewise, consider No. 11, for which
\bea
     \Ln = 2A_4+2D_5~.
\eea
The two $A_4$ factors each contribute two packets of size four, while the two $D_5$ factors each contribute two packets of size five.  The multiplicity of any one distinct eigenvalue must therefore have the form $ m=4a+5b$ for $a,b\in\ZZ$. For example, multiplicity $9=4+5$ is allowed, whereas multiplicity $2$ is impossible.  More strongly, all multiplicities in the complete spectrum must jointly exhaust the eight packets.  Thus e.g.
\bea
    (4,4,4,4,5,5,5,5)~,\hspace{0.3 in}(8,8,10,10)~, \hspace{0.3 in} (18,18)~, \hspace{0.3 in}  (36)~, 
\eea
are each allowed  patterns.  Looking at the results in Table \ref{tab:hessian8d}, we see that the numerical computations indeed realize the second of these.

\section{Details on results}
\label{app:8d}

In this appendix we collect the eight- and nine-dimensional results of our analysis. The nine-dimensional results reproduce the results in \cite{Fraiman:2023cpa}, while the eight-dimensional results are new. 
Due to the large number of eight-dimensional results, below we write only the tachyon-free cases; the complete set of data is attached as an ancillary file. 

\subsection{Nine-dimensional results}

We begin with the nine-dimensional results of our analysis. 
Starting with a parent theory of gauge algebra $\Ls$, we  perform an orbifold specified by the data $\hat w$, $\hat n$, and $\delta_{16}$. This gives a non-supersymmetric theory with algebra $\Ln$, with $N_\mathrm{tach}$ tachyons and $N_0^s$ massless scalars. The theories obtained in this way match precisely with those given in \cite{Fraiman:2023cpa} coming from direct toroidal compactification of ten-dimensional non-supersymmetric heterotic strings, as long as we restrict to the maximally semisimple cases.

\begingroup

\footnotesize
\setlength{\tabcolsep}{3pt}
\renewcommand{\arraystretch}{1.08}
\setlength\LTleft{\fill}
\setlength\LTright{\fill}

\begin{longtable}{|c!{\vrule width 1.2pt}c||c|c|c||c|c|c|}

\caption{The nine-dimensional results of our analysis.}
\label{tab:results9d}
\\
\hline
\# & $L_{\mathrm{SUSY}}$
& $\hat{w}$
& $\hat{n}$
& $\delta_{16}$
& $\Ln$
& $N_{\mathrm{tach}}$
& $N_{0}^{\mathrm{s}}$ \\
\hline\hline
\endfirsthead

\multicolumn{8}{c}{%
\tablename\ \thetable\ continued from the previous page
} \\
\hline
\# & $L_{\mathrm{SUSY}}$
& $\hat{w}$
& $\hat{n}$
& $\delta_{16}$
& $\Ln$
& $N_{\mathrm{tach}}$
& $N_{0}^{\mathrm{s}}$ \\
\hline\hline
\endhead

\hline
\endfoot

\hline
\endlastfoot

1 & \(A_1 + 2E_8\) & \(0\) & \(0\)
& \(\frac{1}{2}(0^4,1^4,0^8)\)
& \(A_1 + D_8 + E_8\) & 80 & 0 \\
\hline

2 & \(A_1 + 2E_8\) & \(1\) & \(-1\) & \(\frac{1}{4}(1^8,0^8)\) & \(2A_1 + E_7 + E_8\) & 116 & 0\\
\hline

3 & \(A_1 + 2E_8\) & \(0\) & \(0\) & \(\frac{1}{2}(0^4,1^4,0^4,1^4)\) & \(A_1 + 2D_8\) & 0 & 0\\
\hline

4 & \(A_1 + 2E_8\) & \(1\) & \(-1\) & \(\frac{1}{4}(1^8,0^4,2^4)\) & \(2A_1 + D_8 + E_7\) & 64 & 0\\ \hline

5 & \(A_1 + 2E_8\) & \(0\) & \(0\) & \(\frac{1}{4}(1^{16})\) & \(3A_1 + 2E_7\) & 20 & 0\\ \hline
6 & \(A_2 + E_7 + E_8\) & \(0\) & \(0\) & \(\frac{1}{2}(0^4,1^4,0^8)\) & \(A_1 + A_2 + D_6 + E_8\) & 96 & 0\\ \hline
7 & \(A_2 + E_7 + E_8\) & \(0\) & \(0\) & \(\frac{1}{2}(0^{12},1^4)\) & \(A_2 + D_8 + E_7\) & 112 & 0\\ \hline
8 & \(A_2 + E_7 + E_8\) & \(0\) & \(0\) & \(\frac{1}{2}(0^4,1^4,0^4,1^4)\) & \(A_1 + A_2 + D_6 + D_8\) & 0 & 0\\ \hline
9 & \(A_2 + E_7 + E_8\) & \(0\) & \(2\) & \(\frac{1}{4}(0^6,-2,2,1^8)\) & \(A_1 + A_2 + 2E_7\) & 12 & 0\\ \hline
10 & \(A_2 + E_7 + E_8\) & \(1\) & \(-1\) & \(\frac{1}{4}(-1,1^5,3,1^9)\) & \(A_1 + A_2 + A_7 + E_7\) & 32 & 0\\ \hline
11 & \(A_2 + E_7 + E_8\) & \(0\) & \(2\) & \(\frac{1}{4}(0^4,2^2,0,4,1^8)\) & \(2A_1 + A_2 + D_6 + E_7\) & 28 & 0\\ \hline
12 & \(A_3 + E_6 + E_8\) & \(1\) & \(1\) & \(\frac{1}{2}(0^5,-1,1,0^9)\) & \(A_3 + E_6 + E_8\) & 60 & 0\\ \hline
13 & \(A_3 + E_6 + E_8\) & \(0\) & \(0\) & \(\frac{1}{2}(0^{12},1^4)\) & \(A_3 + D_8 + E_6\) & 144 & 0\\ \hline
14 & \(A_3 + E_6 + E_8\) & \(1\) & \(1\) & \(\frac{1}{2}(0^5,-1,1,0^5,1^4)\) & \(A_3 + D_8 + E_6\) & 128 & 0\\ \hline
15 & \(A_3 + E_6 + E_8\) & \(2\) & \(-2\) & \(\frac{1}{4}(-1,1^4,3^2,-1,1^8)\) & \(2A_1 + A_3 + A_5 + E_7\) & 36 & 0\\ \hline
16 & \(A_3 + E_6 + E_8\) & \(1\) & \(1\) & \(\frac{1}{4}(-1,1^6,3,1^8)\) & \(2A_1 + A_3 + A_5 + E_7\) & 56 & 0\\ \hline
17 & \(D_9 + E_8\) & \(0\) & \(2\) & \(\frac{1}{2}(0^8,-1,1^7)\) & \(D_9 + E_8\) & 18 & 768\\ \hline
18 & \(D_9 + E_8\) & \(0\) & \(1\) & \(\frac{1}{2}(0^{12},1^4)\) & \(D_4 + D_5 + E_8\) & 40 & 160\\ \hline
19 & \(D_9 + E_8\) & \(0\) & \(0\) & \(\frac{1}{2}(0^4,1^4,0^8)\) & \(D_8 + D_9\) & 16 & 576\\ \hline
20 & \(D_9 + E_8\) & \(0\) & \(2\) & \(\frac{1}{2}(0^4,1^4,-1,1^7)\) & \(D_8 + D_9\) & 0 & 256\\ \hline
21 & \(D_9 + E_8\) & \(0\) & \(1\) & \(\frac{1}{2}(0^4,1^4,0^4,1^4)\) & \(D_4 + D_5 + D_8\) & 0 & 256\\ \hline
22 & \(D_9 + E_8\) & \(0\) & \(1\) & \(\frac{1}{4}(1^8,0^2,2^6)\) & \(3A_1 + D_7 + E_7\) & 4 & 336\\ \hline
23 & \(D_9 + E_8\) & \(1\) & \(0\) & \(\frac{1}{4}(1^8,2,0^2,2^5)\) & \(A_1 + A_3 + D_6 + E_7\) & 16 & 128\\ \hline
24 & \(A_4 + D_5 + E_8\) & \(2\) & \(0\) & \(\frac{1}{2}(0^4,-1,1^2,-1,0^8)\) & \(A_4 + D_5 + E_8\) & 130 & 0\\ \hline
25 & \(A_4 + D_5 + E_8\) & \(0\) & \(0\) & \(\frac{1}{2}(0^{12},1^4)\) & \(A_4 + D_5 + D_8\) & 176 & 0\\ \hline
26 & \(A_4 + D_5 + E_8\) & \(2\) & \(0\) & \(\frac{1}{2}(0^4,-1,1^2,-1,0^4,1^4)\) & \(A_4 + D_5 + D_8\) & 0 & 0\\ \hline
27 & \(A_4 + D_5 + E_8\) & \(2\) & \(-2\) & \(\frac{1}{4}(0^2,2^5,-2,1^8)\) & \(3A_1 + A_3 + A_4 + E_7\) & 44 & 0\\ \hline
28 & \(A_9 + E_8\) & \(0\) & \(0\) & \(\frac{1}{2}(0^4,1^4,0^8)\) & \(A_9 + D_8\) & 336 & 0\\ \hline
29 & \(A_9 + E_8\) & \(1\) & \(1\) & \(\frac{1}{4}(1^8,-1,1^6,3)\) & \(A_1 + A_9 + E_7\) & 40 & 0\\ \hline
30 & \(A_1 + A_8 + E_8\) & \(0\) & \(0\) & \(\frac{1}{2}(0^{12},1^4)\) & \(A_1 + A_8 + D_8\) & 304 & 0\\ \hline
31 & \(A_1 + A_8 + E_8\) & \(0\) & \(0\) & \(\frac{1}{4}(1^{16})\) & \(2A_1 + A_8 + E_7\) & 76 & 0\\ \hline
32 & \(A_1 + A_2 + A_6 + E_8\) & \(0\) & \(0\) & \(\frac{1}{2}(0^{12},1^4)\) & \(A_1 + A_2 + A_6 + D_8\) & 240 & 0\\ \hline
33 & \(A_1 + A_2 + A_6 + E_8\) & \(0\) & \(0\) & \(\frac{1}{4}(-2,2,0^6,1^8)\) & \(2A_1 + A_2 + A_6 + E_7\) & 60 & 0\\ \hline
34 & \(A_4 + A_5 + E_8\) & \(0\) & \(0\) & \(\frac{1}{2}(0^{12},1^4)\) & \(A_4 + A_5 + D_8\) & 208 & 0\\ \hline
35 & \(A_4 + A_5 + E_8\) & \(3\) & \(-3\) & \(\frac{1}{4}(0^2,4,2^4,-4,1^8)\) & \(A_1 + A_4 + A_5 + E_7\) & 80 & 0\\ \hline
36 & \(A_3 + 2E_7\) & \(0\) & \(2\) & \(\frac{1}{2}(0^6,-1,1,0^2,-1,1^5)\) & \(A_3 + 2E_7\) & 6 & 528\\ \hline
37 & \(A_3 + 2E_7\) & \(0\) & \(0\) & \(\frac{1}{2}(0^{12},1^4)\) & \(A_1 + A_3 + D_6 + E_7\) & 28 & 368\\ \hline
38 & \(A_3 + 2E_7\) & \(0\) & \(0\) & \(\frac{1}{2}(0^4,1^4,0^4,1^4)\) & \(2A_1 + A_3 + 2D_6\) & 0 & 256\\ \hline
39 & \(A_3 + 2E_7\) & \(0\) & \(2\) & \(\frac{1}{4}(0^4,2^2,0,4,1^8)\) & \(2A_1 + A_3 + 2D_6\) & 4 & 336\\ \hline
40 & \(A_4 + E_6 + E_7\) & \(0\) & \(0\) & \(\frac{1}{2}(0^{12},1^4)\) & \(A_1 + A_4 + D_6 + E_6\) & 140 & 0\\ \hline
41 & \(A_4 + E_6 + E_7\) & \(2\) & \(-2\) & \(\frac{1}{4}(-1,1^4,3^2,-1,2,-2,0^2,2^4)\) & \(2A_1 + A_4 + A_5 + D_6\) & 148 & 0\\ \hline
42 & \(A_4 + E_6 + E_7\) & \(3\) & \(-3\) & \(\frac{1}{4}(0^5,2^2,-4,3,-3,1^5,3)\) & \(A_4 + A_7 + E_6\) & 80 & 0\\ \hline
43 & \(D_{10} + E_7\) & \(0\) & \(0\) & \(\frac{1}{2}(0^4,1^4,0^8)\) & \(A_1 + D_6 + D_{10}\) & 92 & 0\\ \hline
44 & \(D_{10} + E_7\) & \(1\) & \(0\) & \(\frac{1}{2}(0^8,1,0^4,1^3)\) & \(D_4 + D_6 + E_7\) & 72 & 0\\ \hline
45 & \(D_{10} + E_7\) & \(1\) & \(0\) & \(\frac{1}{2}(0^4,1^5,0^4,1^3)\) & \(A_1 + D_4 + 2D_6\) & 32 & 0\\ \hline
46 & \(D_{10} + E_7\) & \(0\) & \(1\) & \(\frac{1}{4}(-1,1^6,3,0^4,2^4)\) & \(A_7 + 2D_5\) & 0 & 0\\ \hline
47 & \(D_{10} + E_7\) & \(0\) & \(2\) & \(\frac{1}{2}(0^8,-1,1^7)\) & \(D_{10} + E_7\) & 132 & 0\\ \hline
48 & \(D_{10} + E_7\) & \(0\) & \(2\) & \(\frac{1}{2}(0^4,1^4,-1,1^7)\) & \(A_1 + D_6 + D_{10}\) & 64 & 0\\ \hline
49 & \(D_{10} + E_7\) & \(0\) & \(1\) & \(\frac{1}{4}(1^8,0^2,2^6)\) & \(3A_1 + D_6 + D_8\) & 52 & 0\\ \hline
50 & \(A_5 + D_5 + E_7\) & \(0\) & \(2\) & \(\frac{1}{2}(0^4,(-1)^2,1^2,0^8)\) & \(A_5 + D_5 + E_7\) & 40 & 0\\ \hline
51 & \(A_5 + D_5 + E_7\) & \(0\) & \(0\) & \(\frac{1}{2}(0^{12},1^4)\) & \(A_1 + A_5 + D_5 + D_6\) & 100 & 0\\ \hline
52 & \(A_5 + D_5 + E_7\) & \(0\) & \(2\) & \(\frac{1}{2}(0^4,(-1)^2,1^2,0^4,1^4)\) & \(A_1 + A_5 + D_5 + D_6\) & 64 & 0\\ \hline
53 & \(A_5 + D_5 + E_7\) & \(2\) & \(-2\) & \(\frac{1}{2}(0^2,1^5,-1,1,-1,0^2,1^4)\) & \(3A_1 + A_3 + A_5 + D_6\) & 100 & 0\\ \hline
54 & \(A_{10} + E_7\) & \(0\) & \(0\) & \(\frac{1}{2}(0^4,1^4,0^8)\) & \(A_1 + A_{10} + D_6\) & 56 & 0\\ \hline
55 & \(A_1 + A_9 + E_7\) & \(0\) & \(0\) & \(\frac{1}{2}(0^{12},1^4)\) & \(2A_1 + A_9 + D_6\) & 132 & 0\\ \hline
56 & \(A_1 + A_9 + E_7\) & \(2\) & \(0\) & \(\frac{1}{4}(-3,1^6,-3,2,-2,0^2,2^4)\) & \(2A_1 + A_9 + D_6\) & 244 & 0\\ \hline
57 & \(A_1 + A_2 + A_7 + E_7\) & \(0\) & \(0\) & \(\frac{1}{2}(0^{12},1^4)\) & \(2A_1 + A_2 + A_7 + D_6\) & 68 & 0\\ \hline
58 & \(A_1 + A_2 + A_7 + E_7\) & \(2\) & \(-2\) & \(\frac{1}{2}(0^2,1^5,-1,1,-1,0^2,1^4)\) & \(2A_1 + A_2 + A_7 + D_6\) & 76 & 0\\ \hline
59 & \(A_4 + A_6 + E_7\) & \(0\) & \(0\) & \(\frac{1}{2}(0^{12},1^4)\) & \(A_1 + A_4 + A_6 + D_6\) & 80 & 0\\ \hline

60 & \(A_5 + 2E_6\) & \(1\) & \(1\) & \(\frac{1}{2}(0^3,1^4,0^2,-1,1,0^5)\) & \(A_1 + 2A_5 + E_6\) & 72 & 0\\ \hline

61 & \(A_5 + 2E_6\) & \(2\) & \(0\) & \(\frac{1}{4}(-1,1^4,3^2,-1,2,-2,2,0^2,2^3)\) & \(2A_1 + 3A_5\) & 76 & 0\\ \hline

62 & \(D_{11} + E_6\) & \(2\) & \(-2\) & \((0^5,1^2,0,1,0^7)\) & \(D_{11} + E_6\) & 76 & 0\\ \hline

63 & \(D_{11} + E_6\) & \(0\) & \(0\) & \(\frac{1}{2}(0^{12},1^4)\) & \(D_4 + D_7 + E_6\) & 136 & 0\\ \hline

64 & \(D_{11} + E_6\) & \(2\) & \(-2\) & \(\frac{1}{4}(-1,1^4,3^2,-1,4,0^5,2^2)\) & \(3A_1 + A_5 + D_9\) & 28 & 0\\ \hline

65 & \(A_6 + D_5 + E_6\) & \(6\) & \(-6\) & \(\frac{1}{2}(-1,1^3,2^3,-4,4,(-2)^2,0^5)\) & \(A_6 + D_5 + E_6\) & 52 & 0\\ \hline

66 & \(A_6 + D_5 + E_6\) & \(10\) & \(-10\) & \(\frac{1}{2}(0^2,1^2,3^3,-7,7,(-3)^2,0^2,1^3)\) & \(3A_1 + A_3 + A_5 + A_6\) & 52 & 0\\ \hline

67 & \(A_{11} + E_6\) & \(1\) & \(1\) & \(\frac{1}{4}(1,(-1)^3,3^3,1,2,0^2,-2,2^4)\) & \(A_{11} + E_6\) & 0 & 408\\ \hline

68 & \(A_1 + A_{10} + E_6\) & \(4\) & \(-4\) & \(\frac{1}{4}(1^7,-7,6,(-2)^2,0^2,2^3)\) & \(2A_1 + A_5 + A_{10}\) & 136 & 0\\ \hline

69 & \(A_1 + A_2 + A_8 + E_6\) & \(4\) & \(-4\) & \(\frac{1}{2}(0,2,1^5,-3,3,(-1)^2,0^2,1^3)\) & \(2A_1 + A_2 + A_5 + A_8\) & 40 & 0\\ \hline

70 & \(D_{17}\) & \(2\) & \(-2\) & \(\frac{1}{2}(0^4,1^3,-1,3,1^7)\) & \(D_4 + D_{13}\) & 8 & 416\\ \hline

71 & \(D_{17}\) & \(4\) & \(-4\) & \(\frac{1}{4}(-1,1^4,3^2,-5,10,2^7)\) & \(D_5 + D_{12}\) & 32 & 0\\ \hline

72 & \(D_{17}\) & \(0\) & \(0\) & \((0^{15},1)\) & \(D_{17}\) & 34 & 1088\\ \hline

73 & \(A_1 + D_{16}\) & \(2\) & \(-2\) & \(\frac{1}{2}(-1,1^6,-1,3,1^7)\) & \(A_1 + D_{16}\) & 160 & 0\\ \hline

74 & \(A_1 + D_{16}\) & \(2\) & \(-2\) & \(\frac{1}{4}((-1)^2,1^5,-3,6,2^7)\) & \(3A_1 + D_{14}\) & 116 & 0\\ \hline

75 & \(A_1 + D_{16}\) & \(0\) & \(0\) & \(\frac{1}{2}(0^4,1^{12})\) & \(A_1 + D_4 + D_{12}\) & 40 & 0\\ \hline

76 & \(A_1 + A_2 + D_{14}\) & \(2\) & \(-2\) & \(\frac{1}{2}(-1,1^6,-1,2,0^7)\) & \(A_1 + A_2 + D_{14}\) & 208 & 0\\ \hline

77 & \(A_1 + A_2 + D_{14}\) & \(4\) & \(-4\) & \(\frac{1}{2}(0^4,1^3,-3,5,1^7)\) & \(3A_1 + A_2 + D_{12}\) & 40 & 0\\ \hline

78 & \(A_1 + A_2 + D_{14}\) & \(2\) & \(-2\) & \(\frac{1}{2}(-1,1,0^4,1,-1,3,1^7)\) & \(A_1 + A_2 + D_4 + D_{10}\) & 56 & 0\\ \hline

79 & \(A_4 + D_{13}\) & \(2\) & \(-2\) & \(\frac{1}{2}(-1,1^6,-1,2,0^7)\) & \(A_4 + D_{13}\) & 306 & 0\\ \hline

80 & \(A_4 + D_{13}\) & \(7\) & \(-6\) & \(\frac{1}{2}(0^3,1^3,2,-5,7,1^7)\) & \(A_4 + D_4 + D_9\) & 88 & 0\\ \hline

81 & \(D_5 + D_{12}\) & \(2\) & \(-2\) & \(\frac{1}{2}(-1,1^6,-1,2,0^7)\) & \(D_5 + D_{12}\) & 56 & 480\\ \hline

82 & \(D_5 + D_{12}\) & \(0\) & \(0\) & \(\frac{1}{2}(0^4,1^4,0^8)\) & \(D_5 + D_{12}\) & 10 & 608\\ \hline

83 & \(D_5 + D_{12}\) & \(0\) & \(1\) & \(\frac{1}{2}(-1,1^3,0^2,1^2,0^2,1^6)\) & \(D_4 + D_5 + D_8\) & 8 & 416\\ \hline

84 & \(D_5 + D_{12}\) & \(1\) & \(0\) & \(\frac{1}{2}(0^2,1^2,0,1^2,0,1^8)\) & \(4A_1 + A_3 + D_{10}\) & 20 & 208\\ \hline

85 & \(A_{12} + D_5\) & \(6\) & \(-6\) & \(\frac{1}{2}(-1,1^3,2^3,-4,5,(-1)^6,1)\) & \(A_{12} + D_5\) & 166 & 0\\ \hline

86 & \(A_1 + A_{11} + D_5\) & \(0\) & \(0\) & \(\frac{1}{2}(0^{12},1^4)\) & \(A_1 + A_{11} + D_5\) & 142 & 0\\ \hline

87 & \(A_1 + A_{11} + D_5\) & \(0\) & \(0\) & \(\frac{1}{4}(1^{16})\) & \(3A_1 + A_3 + A_{11}\) & 100 & 0\\ \hline

88 & \(A_1 + A_2 + A_9 + D_5\) & \(4\) & \(-4\) & \(\frac{1}{2}(1^7,-3,3,(-1)^3,0^4)\) & \(A_1 + A_2 + A_9 + D_5\) & 100 & 0\\ \hline

89 & \(A_1 + A_2 + A_9 + D_5\) & \(6\) & \(-6\) & \(\frac{1}{2}(0^2,1^5,-5,5,(-1)^3,0^2,1^2)\) & \(3A_1 + A_2 + A_3 + A_9\) & 28 & 0\\ \hline

90 & \(A_4 + A_8 + D_5\) & \(12\) & \(-12\) & \(\frac{1}{2}(-1,1^2,3^4,-9,9,(-3)^3,0^4)\) & \(A_4 + A_8 + D_5\) & 82 & 0\\ \hline

91 & \(A_7 + 2D_5\) & \(0\) & \(0\) & \(\frac{1}{2}(0^4,1^4,0^8)\) & \(A_7 + 2D_5\) & 66 & 0\\ \hline

92 & \(A_7 + 2D_5\) & \(2\) & \(-2\) & \(\frac{1}{2}(0^2,1^5,-1,2,0^5,1^2)\) & \(4A_1 + 2A_3 + A_7\) & 36 & 0\\ \hline

93 & \(2A_1 + A_{15}\) & \(0\) & \(0\) & \(\frac{1}{4}(1^{16})\) & \(2A_1 + A_{15}\) & 4 & 488\\ \hline

94 & \(2A_1 + A_2 + A_{13}\) & \(0\) & \(0\) & \(\frac{1}{4}(1^8,0^6,-2,2)\) & \(2A_1 + A_2 + A_{13}\) & 88 & 0\\ \hline

95 & \(2A_1 + 2A_2 + A_{11}\) & \(0\) & \(0\) & \(\frac{1}{2}(-1,1,0^{12},-1,1)\) & \(2A_1 + 2A_2 + A_{11}\) & 22 & 0\\\hline

\end{longtable}

\endgroup

\subsection{Eight-dimensional results}

The eight-dimensional results are given in Table \ref{tab:expanded-8d-data} below. Starting with the maximally enhanced supersymmetric theory labelled by $k$ in the notation of \cite[Table 12]{Font:2020rsk}, which has algebra $\Ls$, 
we perform an orbifold specified by the data $(\hat w^1,\hat w^2;\hat n_1,\hat n_2)$ and $\delta_{16}$. 
Given the parent moduli $(G,B,A_i)$, the full left- and right-moving shift is reconstructed by substituting these data into (\ref{eq:delta16wn}).  Every listed representative obeys
\bea
\label{eq:appendix-level-matching}
2\delta_{16} \in\Gamma_{16}~, \hspace{0.4 in} \delta^2=\delta_{16}^2+\half (\hat w^1 \hat n_1 +\hat w^2 \hat n_2)\in\mathbb Z~.
\eea
The integers $\hat w^i$ and $\hat n_i$ have not always been reduced to zero or one.  This is intentional: adding even integers is accompanied by a compensating Narain-lattice translation and a change of $\delta_{16}$, so an unreduced representative is often shorter or more transparent.  Only the equivalence class of the complete vector $\delta$ is physical.

The result of the orbifold is a non-supersymmetric theory with left-moving algebra $\Ln$.  As in the main table, the two universal right-moving factors $\mathfrak u(1)^2_R$ are suppressed, though when one of these right-moving $\mathfrak u(1)$ factors is enhanced to $\mathfrak{su}(2)$ we indicate this by appending $A_1^{(\mathrm R)}$ to $\Ln$.
No row contains an additional left-moving $\mathfrak u(1)$.  The quantities $N_0^{\rm s}$, $N_0^{\rm f}$, and $\Lambda^{(8)}$ use the conventions of Section \ref{sec:light-states}.  In particular, $N_0^{\rm s}$ counts shifted-sector $O_8$ knife-edge states, not the universal dilaton or neutral moduli.

As an illustration, entry~22 is based on parent $k=169$ and uses a shift vector with $(\hat w^1,\hat w^2;\hat n_1,\hat n_2)=(0,0;0,2)$.  Combining this tuple with the displayed $\delta_{16}$ and the parent moduli reconstructs the full shift.  Its invariant left-moving algebra is $2A_1+2D_8$, it has $N_0^{\rm s}=1024$ and $N_0^{\rm f}=256$, and its one-loop cosmological constant is $\Lambda^{(8)}=-1.87 \times 10^{-6}$.  Entry~23 starts from the same parent but uses a different shift and has a different invariant algebra and spectrum.  This illustrates why neither the parent label nor the abstract gauge algebra alone uniquely specifies a row.

\begingroup
\tiny
\setlength{\tabcolsep}{2pt}
\renewcommand{\arraystretch}{1.0}

\setlength\LTleft{\fill}
\setlength\LTright{\fill}

\begin{longtable}{|c!{\vrule width 1.2pt}c|c||c|c||c|c|c|c|}
\caption{Expanded form of the tachyon-free eight-dimensional entries listed in Table \ref{tab:8d}.}
\label{tab:expanded-8d-data}
\\
\hline
\# & $k$ & $L_{\rm SUSY}$ & $(\hat w^1,\hat w^2;\hat n_1,\hat n_2)$ & $\delta_{16}$
& $\Ln$ & $N_0^{\rm s}$ & $N_0^{\rm f}$ & $\Lambda^{(8)}\times 10^6$ \\
\hline\hline
\endfirsthead

\multicolumn{9}{c}{\tablename\ \thetable\ continued from the previous page} \\
\hline
\# & $k$ & $L_{\rm SUSY}$ & $(\hat w^1,\hat w^2;\hat n_1,\hat n_2)$ & $\delta_{16}$
& $\Ln$ & $N_0^{\rm s}$ & $N_0^{\rm f}$ & $\Lambda^{(8)}\times 10^6$ \\
\hline\hline
\endhead

\hline
\endfoot

\hline
\endlastfoot
1 & $1$ & $6A_{3}$ & $(0,0;2,1)$ & $\frac{1}{2}(-1,1,0^6,-1,1,(-1)^2,1^2,0^2)$ & $6A_{3}+A_1^{(\rm R)}$ & $360$ & $0$ & $63.20$ \\ \hline
2 & $1$ & $6A_{3}$ & $(0,0;0,0)$ & $\frac{1}{2}(1^2,0^4,1^4,0^4,1^2)$ & $6A_{3}$ & $0$ & $108$ & $106.82$ \\ \hline
3 & $13$ & $3A_{6}$ & $(0,1;0,-1)$ & $\frac{1}{4}((-1)^5,3,1^3,(-1)^2,3^4,1)$ & $3A_{6}+A_1^{(\rm R)}$ & $252$ & $0$ & $67.41$ \\ \hline
4 & $25$ & $4A_{1} + 2A_{7}$ & $(0,0;0,0)$ & $\frac{1}{2}(-1,1,-1,1,0^8,-1,1,-1,1)$ & $4A_{1} + 2A_{7}$ & $0$ & $156$ & $106.82$ \\ \hline
5 & $28$ & $2A_{1} + 3A_{3} + A_{7}$ & $(0,0;0,0)$ & $\frac{1}{2}(0^4,1^4,0^4,(-1)^2,1^2)$ & $2A_{1} + 3A_{3} + A_{7}$ & $0$ & $130$ & $106.30$ \\ \hline
6 & $54(1)$ & $2A_{9}$ & $(3,1;-3,-1)$ & $\frac{1}{4}(-1,1^6,-5,5,(-1)^6,1)$ & $2A_{9}+2A_1^{(\rm R)}$ & $720$ & $0$ & $19.59$ \\ \hline
7 & $83(1)$ & $2A_{2} + A_{3} + A_{11}$ & $(1,0;1,0)$ & $\frac{1}{4}(1,(-1)^3,3^3,1,2,0^2,-2,2^4)$ & $2A_{2} + A_{3} + A_{11}+A_1^{(\rm R)}$ & $312$ & $0$ & $66.09$ \\ \hline
8 & $88$ & $A_{2} + A_{5} + A_{11}$ & $(1,0;1,0)$ & $\frac{1}{4}(1,(-1)^3,3^3,1,2,0^2,-2,2^4)$ & $A_{2} + A_{5} + A_{11}+A_1^{(\rm R)}$ & $336$ & $0$ & $71.57$ \\ \hline
9 & $89$ & $A_{1} + A_{6} + A_{11}$ & $(1,4;1,-4)$ & $\frac{1}{4}(1,(-1)^3,3^3,1,4,2^2,(-4)^2,0^3)$ & $A_{1} + A_{6} + A_{11}+A_1^{(\rm R)}$ & $352$ & $0$ & $61.36$ \\ \hline
10 & $111(2)$ & $A_{1} + A_{17}$ & $(1,1;1,-1)$ & $\frac{1}{4}(1,(-1)^3,3^3,1,3,1^6,-1)$ & $A_{1} + A_{17}+2A_1^{(\rm R)}$ & $1232$ & $0$ & $-87.21$ \\ \hline
11 & $113$ & $2A_{4} + 2D_{5}$ & $(0,0;0,0)$ & $\frac{1}{2}(0^4,1^4,0^4,1^4)$ & $2A_{4} + 2D_{5}$ & $0$ & $100$ & $145.97$ \\ \hline
12 & $121$ & $A_{1} + A_{7} + 2D_{5}$ & $(0,0;0,0)$ & $\frac{1}{2}(0^4,1^4,0^4,1^4)$ & $A_{1} + A_{7} + 2D_{5}$ & $0$ & $170$ & $147.45$ \\ \hline
13 & $122$ & $A_{1} + A_{2} + A_{3} + A_{7} + D_{5}$ & $(0,0;0,0)$ & $\frac{1}{2}(0^4,1^4,0^4,1^4)$ & $A_{1} + A_{2} + A_{3} + A_{7} + D_{5}$ & $0$ & $130$ & $130.69$ \\ \hline
14 & $123$ & $2A_{1} + A_{4} + A_{7} + D_{5}$ & $(0,0;0,0)$ & $\frac{1}{2}(-1,1,-1,1,0^8,1^4)$ & $2A_{1} + A_{4} + A_{7} + D_{5}$ & $0$ & $110$ & $120.48$ \\ \hline
15 & $124$ & $A_{8} + 2D_{5}$ & $(0,1;0,-1)$ & $\frac{1}{2}(1^2,2,0^9,1^4)$ & $A_{8} + 2D_{5}$ & $0$ & $100$ & $111.94$ \\ \hline
16 & $136$ & $3D_{6}$ & $(0,0;2,0)$ & $\frac{1}{2}(0^6,-1,1^3,0^5,2)$ & $3D_{6}+2A_1^{(\rm R)}$ & $720$ & $0$ & $19.59$ \\ \hline
17 & $137$ & $2A_{3} + 2D_{6}$ & $(0,0;2,0)$ & $\frac{1}{2}(0^7,2,0^4,1^4)$ & $2A_{1} + 2A_{3} + D_{4} + D_{6}$ & $448$ & $128$ & $90.05$ \\ \hline
18 & $137$ & $2A_{3} + 2D_{6}$ & $(0,0;0,0)$ & $(0^5,1,0^9,1)$ & $2A_{3} + 2D_{6}$ & $0$ & $180$ & $106.82$ \\ \hline
19 & $137$ & $2A_{3} + 2D_{6}$ & $(0,0;0,0)$ & $\frac{1}{2}(0^2,1^4,0^6,1^4)$ & $4A_{1} + 2A_{3} + 2D_{4}$ & $256$ & $176$ & $105.88$ \\ \hline
20 & $154$ & $A_{7} + D_{5} + D_{6}$ & $(0,0;0,0)$ & $\frac{1}{2}(0^4,1^4,0^4,1^4)$ & $A_{7} + D_{5} + D_{6}$ & $0$ & $190$ & $106.30$ \\ \hline
21 & $167$ & $A_{1} + A_{5} + D_{5} + D_{7}$ & $(2,0;0,1)$ & $\frac{1}{4}(0^4,-2,2^2,-2,3,-1,1^6)$ & $A_{1} + A_{3} + A_{5} + D_{4} + D_{5}$ & $416$ & $128$ & $96.83$ \\ \hline
22 & $169$ & $2A_{1} + 2D_{8}$ & $(0,0;0,2)$ & $\frac{1}{2}(0^3,2,-1,1^3,0^4,1^4)$ & $2A_{1} + 2D_{8}$ & $1024$ & $256$ & $-1.87$ \\ \hline
23 & $169$ & $2A_{1} + 2D_{8}$ & $(0,0;1,1)$ & $\frac{1}{4}(1^{16})$ & $2A_{1} + 4D_{4}$ & $512$ & $256$ & $104.94$ \\ \hline
24 & $177$ & $2D_{5} + D_{8}$ & $(0,0;0,0)$ & $\frac{1}{2}(0^4,1^4,0^4,1^4)$ & $2D_{5} + D_{8}$ & $0$ & $228$ & $106.82$ \\ \hline
25 & $177$ & $2D_{5} + D_{8}$ & $(0,1;0,-1)$ & $\frac{1}{2}(0^4,1^5,0^2,1,0^3,2)$ & $2D_{5} + D_{8}$ & $640$ & $128$ & $74.22$ \\ \hline
26 & $177$ & $2D_{5} + D_{8}$ & $(0,1;1,-1)$ & $\frac{1}{2}(-1,1^3,0^2,1^3,0^2,1,0^2,1^2)$ & $4A_{1} + A_{3} + D_{5} + D_{6}$ & $352$ & $176$ & $97.96$ \\ \hline
27 & $177$ & $2D_{5} + D_{8}$ & $(0,0;1,0)$ & $(0^{16})$ & $2D_{4} + 2D_{5}$ & $256$ & $224$ & $105.88$ \\ \hline
28 & $179$ & $2D_{9}$ & $(0,0;2,2)$ & $\frac{1}{2}(-1,1^7,-1,1^7)$ & $2D_{9}$ & $0$ & $324$ & $106.82$ \\ \hline
29 & $179$ & $2D_{9}$ & $(0,0;2,1)$ & $\frac{1}{2}(-1,1^7,0^4,1^4)$ & $D_{4} + D_{5} + D_{9}$ & $448$ & $224$ & $90.05$ \\ \hline
30 & $187$ & $A_{4} + D_{5} + D_{9}$ & $(0,2;2,0)$ & $\frac{1}{2}(0^4,-1,1^2,(-1)^2,1^7)$ & $A_{4} + D_{5} + D_{9}$ & $0$ & $180$ & $117.13$ \\ \hline
31 & $187$ & $A_{4} + D_{5} + D_{9}$ & $(0,2;1,0)$ & $\frac{1}{2}(0^4,-1,1^2,-1,0^4,1^4)$ & $A_{4} + D_{4} + 2D_{5}$ & $160$ & $160$ & $124.00$ \\ \hline
32 & $188$ & $2A_{1} + 2A_{3} + D_{10}$ & $(2,0;-1,2)$ & $\frac{1}{4}(0^4,2^2,4,0,3^2,(-1)^2,1^4)$ & $2A_{1} + 2A_{3} + D_{4} + D_{6}$ & $256$ & $208$ & $105.88$ \\ \hline
33 & $188$ & $2A_{1} + 2A_{3} + D_{10}$ & $(0,0;2,2)$ & $\frac{1}{2}(0^4,1^4,0^2,1^2,0^2,1^2)$ & $4A_{1} + 2A_{3} + D_{8}$ & $256$ & $240$ & $105.88$ \\ \hline
34 & $299$ & $2A_{1} + 2A_{4} + E_{8}$ & $(0,0;0,0)$ & $\frac{1}{2}(-1,1,-1,1,0^8,1^4)$ & $2A_{1} + 2A_{4} + D_{8}$ & $0$ & $192$ & $142.49$ \\ \hline
35 & $191$ & $3A_{1} + A_{5} + D_{10}$ & $(0,2;1,-2)$ & $\frac{1}{4}(0^2,2^2,(-2)^2,0^2,3,-1,1^6)$ & $3A_{1} + A_{5} + D_{4} + D_{6}$ & $576$ & $192$ & $79.66$ \\ \hline
36 & $191$ & $3A_{1} + A_{5} + D_{10}$ & $(2,2;0,-2)$ & $\frac{1}{2}(-1,1^3,0^3,-2,2,0,-1,1,0^2,1^2)$ & $5A_{1} + A_{5} + D_{8}$ & $384$ & $248$ & $98.28$ \\ \hline
37 & $195$ & $A_{1} + A_{2} + D_{5} + D_{10}$ & $(0,0;1,0)$ & $\frac{1}{4}(-2,2^3,0^2,2^2,1^8)$ & $A_{1} + A_{2} + D_{4} + D_{5} + D_{6}$ & $192$ & $240$ & $131.32$ \\ \hline
38 & $195$ & $A_{1} + A_{2} + D_{5} + D_{10}$ & $(0,0;1,0)$ & $\frac{1}{4}(0^2,2^2,0^2,2^2,1^2,-1,3,1^4)$ & $3A_{1} + A_{2} + A_{3} + D_{4} + D_{6}$ & $192$ & $216$ & $131.32$ \\ \hline
39 & $195$ & $A_{1} + A_{2} + D_{5} + D_{10}$ & $(0,0;2,0)$ & $\frac{1}{2}(0^4,1^4,0^2,1^2,0^2,1^2)$ & $3A_{1} + A_{2} + D_{5} + D_{8}$ & $128$ & $232$ & $129.16$ \\ \hline
40 & $199$ & $A_{1} + A_{2} + A_{4} + D_{11}$ & $(0,0;0,0)$ & $\frac{1}{4}(-2,2,0^2,2^4,1^8)$ & $A_{1} + A_{2} + A_{4} + D_{5} + D_{6}$ & $0$ & $184$ & $155.53$ \\ \hline
41 & $199$ & $A_{1} + A_{2} + A_{4} + D_{11}$ & $(0,0;0,0)$ & $\frac{1}{4}(-2,2,0^6,1^8)$ & $A_{1} + A_{2} + A_{3} + A_{4} + D_{8}$ & $0$ & $224$ & $154.51$ \\ \hline
42 & $256$ & $2A_{2} + 2E_{7}$ & $(0,0;0,0)$ & $\frac{1}{2}(0^4,1^4,0^4,1^4)$ & $2A_{1} + 2A_{2} + 2D_{6}$ & $0$ & $272$ & $166.26$ \\ \hline
43 & $316$ & $2A_{2} + D_{6} + E_{8}$ & $(0,0;0,0)$ & $\frac{1}{2}(0^2,1^4,0^6,1^4)$ & $2A_{1} + 2A_{2} + D_{4} + D_{8}$ & $0$ & $288$ & $166.26$ \\ \hline
44 & $203$ & $A_{1} + A_{2} + A_{3} + D_{12}$ & $(2,0;-1,0)$ & $\frac{1}{2}(0^6,1,-1,2,0,1^6)$ & $A_{1} + A_{2} + A_{3} + D_{4} + D_{8}$ & $256$ & $304$ & $133.48$ \\ \hline
45 & $203$ & $A_{1} + A_{2} + A_{3} + D_{12}$ & $(0,0;0,0)$ & $\frac{1}{4}(1^8,0^2,2^6)$ & $A_{1} + A_{2} + A_{3} + 2D_{6}$ & $128$ & $208$ & $129.16$ \\ \hline
46 & $204$ & $2A_{1} + A_{4} + D_{12}$ & $(2,0;-1,0)$ & $\frac{1}{2}(0^6,1,-1,2,0,1^6)$ & $2A_{1} + A_{4} + D_{4} + D_{8}$ & $256$ & $288$ & $124.78$ \\ \hline
47 & $204$ & $2A_{1} + A_{4} + D_{12}$ & $(2,2;-2,-2)$ & $\frac{1}{2}(0^2,1^2,0^3,-2,2,0,1^6)$ & $2A_{1} + A_{4} + 2D_{6}$ & $256$ & $272$ & $124.78$ \\ \hline
48 & $205$ & $A_{1} + D_{5} + D_{12}$ & $(2,0;-1,0)$ & $\frac{1}{2}(0^6,1,-1,2,0,1^6)$ & $A_{1} + D_{4} + D_{5} + D_{8}$ & $256$ & $336$ & $148.43$ \\ \hline
49 & $205$ & $A_{1} + D_{5} + D_{12}$ & $(2,0;-2,0)$ & $\frac{1}{2}(0^2,1^5,-1,2,0,1^6)$ & $3A_{1} + A_{3} + 2D_{6}$ & $256$ & $296$ & $148.43$ \\ \hline
50 & $206$ & $D_{6} + D_{12}$ & $(2,0;-1,0)$ & $\frac{1}{2}(0^6,1,-1,2,0,1^6)$ & $D_{4} + D_{6} + D_{8}$ & $512$ & $352$ & $104.94$ \\ \hline
51 & $206$ & $D_{6} + D_{12}$ & $(0,1;0,1)$ & $\frac{1}{2}(0^{10},1^6)$ & $2A_{1} + D_{4} + 2D_{6}$ & $512$ & $304$ & $104.94$ \\ \hline
52 & $314$ & $A_{1} + A_{4} + D_{5} + E_{8}$ & $(0,0;0,0)$ & $\frac{1}{2}(0^4,1^4,0^4,1^4)$ & $A_{1} + A_{4} + D_{5} + D_{8}$ & $0$ & $288$ & $175.58$ \\ \hline
53 & $315$ & $A_{5} + D_{5} + E_{8}$ & $(0,0;0,0)$ & $\frac{1}{2}(0^4,1^4,0^4,1^4)$ & $A_{5} + D_{5} + D_{8}$ & $0$ & $288$ & $147.19$ \\ \hline
54 & $209$ & $D_{5} + D_{13}$ & $(0,3;1,-2)$ & $\frac{1}{4}(-1,1,7,(-1)^4,1,0^2,2^6)$ & $2A_{1} + A_{3} + D_{6} + D_{7}$ & $256$ & $256$ & $105.88$ \\ \hline
55 & $209$ & $D_{5} + D_{13}$ & $(5,0;-5,0)$ & $\frac{1}{2}(0^3,1^4,-4,5,0^2,1^5)$ & $2D_{5} + D_{8}$ & $256$ & $288$ & $105.88$ \\ \hline
56 & $316$ & $2A_{2} + D_{6} + E_{8}$ & $(0,0;0,0)$ & $\frac{1}{2}(-1,1^7,0^4,1^4)$ & $2A_{2} + D_{6} + D_{8}$ & $0$ & $320$ & $166.26$ \\ \hline
57 & $211$ & $2A_{1} + A_{2} + D_{14}$ & $(0,0;0,0)$ & $\frac{1}{2}(-1,1,0^8,1^6)$ & $2A_{1} + A_{2} + D_{6} + D_{8}$ & $0$ & $384$ & $187.77$ \\ \hline
58 & $212$ & $A_{1} + A_{3} + D_{14}$ & $(0,0;0,0)$ & $\frac{1}{2}(-1,1,0^8,1^6)$ & $A_{1} + A_{3} + D_{6} + D_{8}$ & $0$ & $384$ & $165.36$ \\ \hline
59 & $213$ & $A_{4} + D_{14}$ & $(0,0;0,0)$ & $\frac{1}{4}(1,(-1)^2,1^{13})$ & $A_{4} + D_{6} + D_{8}$ & $256$ & $320$ & $124.78$ \\ \hline
60 & $214$ & $A_{1} + A_{2} + D_{15}$ & $(0,0;1,0)$ & $\frac{1}{4}(1,(-1)^2,1^{13})$ & $A_{1} + A_{2} + D_{7} + D_{8}$ & $256$ & $352$ & $133.48$ \\ \hline
61 & $214$ & $A_{1} + A_{2} + D_{15}$ & $(4,0;-3,0)$ & $\frac{1}{4}(-1,1^4,3^2,-5,5^2,-5,-1,1^4)$ & $A_{1} + A_{2} + D_{6} + D_{9}$ & $128$ & $280$ & $129.16$ \\ \hline
62 & $296$ & $2A_{1} + 2E_{8}$ & $(0,0;0,0)$ & $\frac{1}{2}(0^4,1^4,0^4,1^4)$ & $2A_{1} + 2D_{8}$ & $0$ & $512$ & $211.74$ \\ \hline
63 & $297$ & $A_{2} + 2E_{8}$ & $(0,0;0,0)$ & $\frac{1}{2}(0^4,1^4,0^4,1^4)$ & $A_{2} + 2D_{8}$ & $0$ & $512$ & $196.54$ \\ \hline
64 & $217$ & $A_{1} + D_{17}$ & $(4,0;-5,0)$ & $\frac{1}{2}(-1,1^6,-3,5,1^7)$ & $A_{1} + D_{8} + D_{9}$ & $256$ & $416$ & $148.43$ \\ \hline
65 & $218$ & $D_{18}$ & $(1,0;0,0)$ & $\frac{1}{4}(-1,1^6,-1,0^2,2^6)$ & $D_{8} + D_{10}$ & $512$ & $448$ & $104.94$ \\ \hline
66 & $241(1)$ & $A_{1} + A_{11} + E_{6}$ & $(1,0;1,0)$ & $\frac{1}{4}(1,(-1)^3,3^3,1,2,0^2,-2,2^4)$ & $A_{1} + A_{11} + E_{6}+A_1^{(\rm R)}$ & $412$ & $0$ & $82.53$ \\ \hline
67 & $242$ & $A_{12} + E_{6}$ & $(3,3;-3,-1)$ & $\frac{1}{4}(-1,1^6,-5,6,0^3,-4,2^3)$ & $A_{12} + E_{6}+A_1^{(\rm R)}$ & $456$ & $0$ & $52.15$ \\ \hline
68 & $255$ & $D_{12} + E_{6}$ & $(2,0;-1,0)$ & $\frac{1}{2}(0^6,1,-1,2,0,1^6)$ & $D_{4} + D_{8} + E_{6}$ & $768$ & $256$ & $61.05$ \\ \hline
69 & $255$ & $D_{12} + E_{6}$ & $(0,3;0,-2)$ & $\frac{1}{4}(-1,1^2,7,(-1)^4,0^2,2^6)$ & $A_{1} + A_{5} + 2D_{6}$ & $384$ & $248$ & $98.28$ \\ \hline
70 & $259$ & $A_{1} + 2A_{3} + A_{4} + E_{7}$ & $(0,0;2,0)$ & $\frac{1}{2}(-1,1,0^4,-1,1,0^4,1^4)$ & $2A_{1} + 2A_{3} + A_{4} + D_{6}$ & $128$ & $160$ & $120.95$ \\ \hline
71 & $290$ & $A_{1} + D_{10} + E_{7}$ & $(0,1;2,-1)$ & $\frac{1}{2}(0^5,2,1^2,-1,1^7)$ & $A_{1} + D_{10} + E_{7}+2A_1^{(\rm R)}$ & $1232$ & $0$ & $-87.21$  \\ \hline

\end{longtable}

\endgroup

\bibliographystyle{ytamsalpha}

\baselineskip=.97\baselineskip

\bibliography{ref}

\end{document}